\documentclass[12pt,a4paper]{article}
\usepackage{latexsym,graphicx,multirow}
\usepackage{float}
\usepackage{amssymb}
\usepackage{amscd}
\usepackage{amsthm}
\usepackage[left=2cm,top=2cm,right=1.8cm,bottom=2cm]{geometry}
\usepackage[hyperfootnotes=false]{hyperref}
\usepackage{epstopdf}
\usepackage{cite}
\usepackage{url}
\usepackage{orcidlink}
\usepackage[utf8]{inputenc}
\usepackage{enumerate}
\usepackage{caption}
\usepackage{subcaption}
\usepackage{amsmath}
\usepackage[mathscr]{euscript}
\usepackage{calrsfs}
\usepackage{framed}
\usepackage{xparse}
\usepackage{fancyvrb}
\usepackage{lscape}
\usepackage{tikz}
\usetikzlibrary{patterns.meta}
\usepackage{multirow}
\usepackage{latexsym,graphicx,multirow}
\theoremstyle{plain}
\usepackage{longtable}
\usetikzlibrary{positioning,shapes}
\usepackage{algorithm}
\usepackage{algpseudocode}
\newtheorem{theorem}{Theorem}[section]
\newtheorem{corollary}{Corrolary}[section]
\newtheorem{proposition}{Proposition}[section]
\newtheorem{remark}{Remark}[section]
\newtheorem{lemma}{Lemma}[section]
\newtheorem{definition}{Definition}[section]

\newcommand{\be}{\begin{equation}}
	\newcommand{\ee}{\end{equation}}
\newcommand{\bp}{\begin{proposition}}
	\newcommand{\ep}{\end{proposition}}
\newcommand{\ben}{\begin{equation*}}
	\newcommand{\een}{\end{equation*}}
\newcommand{\bd}{\begin{definition}}
	\newcommand{\ed}{\end{definition}}
\newcommand{\bl}{\begin{lemma}}
	\newcommand{\el}{\end{lemma}}
\newcommand{\bn}{\begin{notation}}
	\newcommand{\en}{\end{notation}}
\newcommand{\bcon}{\begin{construction}}
	\newcommand{\econ}{\end{construction}}
\newcommand{\bea}{\begin{eqnarray}}
	\newcommand{\eea}{\end{eqnarray}}
\newcommand{\bee}{\begin{eqnarray*}}
	\newcommand{\eee}{\end{eqnarray*}}
\newcommand{\bt}{\begin{theorem}}
	\newcommand{\et}{\end{theorem}}
\newcommand{\br}{\begin{remark}}
	\newcommand{\er}{\end{remark}}
\newcommand{\bo}{\begin{observation}}
	\newcommand{\eo}{\end{observation}}
\newcommand{\bex}{\begin{example}}
	\newcommand{\eex}{\end{example}}
\newcommand{\bc}{\begin{corollary}}
	\newcommand{\ec}{\end{corollary}}
\newcommand{\numsize}[1]{{\fontsize{11pt}{13pt}\selectfont #1}}
\usepackage{array}
\newcolumntype{N}{>{\numsize}c}
\NewDocumentCommand{\INTERVALINNARDS}{ m m }{
	#1 {,} #2
}

\makeatother
\NewDocumentCommand{\interval}{ s m >{\SplitArgument{1}{,}}m m o }{
	\IfBooleanTF{#1}{
		\left#2 \INTERVALINNARDS #3 \right#4
	}{
		\IfValueTF{#5}{
			#5{#2} \INTERVALINNARDS #3 #5{#4}
		}{
			#2 \INTERVALINNARDS #3 #4
		}
	}
}
\usepackage[symbol]{footmisc}
\renewcommand{\thefootnote}{\fnsymbol{footnote}}
\begin{document}	
\begin{center}
	\large{\bf{Relative Variability Estimation for the Power Lindley Distribution with Progressive Type-I Interval Censored Data}} \\
	\vspace{10mm}
	\normalsize{ Bankitdor M. Nongrum$^{1}$\orcidlink{0009-0003-8828-5046}, Adarsha Kumar Jena$^{2}$\footnote[1]{Corresponding author.} \let\thefootnote\relax\footnote{\textit{E-mail address:} jadarsha@gmail.com (Adarsha Kumar Jena)} \orcidlink{0000-0001-8372-8176}
	} \\
	\vspace{5mm}
	\normalsize{$^{1,2}$Department of Mathematics, Saitsohpen, Sohra, East Khasi Hills District, Meghalaya, India - 793108}\\
	\vspace{6mm}
\end{center}
\begin{abstract}
	\noindent The measures of relative variability, such as the coefficient of variation, are highly utilized tools of inference in several interdisciplinary fields such as reliability, finance, quality control, and biological sciences. Inferential studies of these measures are highly limited for lifetime models, such as the flexible power Lindley distribution, particularly under progressive type-I interval censoring.	In this article, frequentist approaches including the midpoint approximation, maximum likelihood estimation, method of moments, bootstrap, and non-linear least squares methods, and Bayesian inference using slice sampling are applied for estimation of these measures of relative variability for the power Lindley distribution under progressive type-I interval censoring. Confidence intervals are also constructed based on asymptotic theory, bootstrap and Bayesian paradigm. A discussion on choosing optimal monitoring intervals is also highlighted. A comprehensive simulation study is conducted to evaluate the performance of the proposed estimators across various censoring plans and sample sizes. A real data application illustrates the practical utility of the proposed methodologies. The results indicate that the Bayesian framework generally exhibits superior performance in both point and interval estimation. 
\end{abstract}
\smallskip 
{\bf Keywords}: Coefficient of Variation; Progressive Type-I Interval Censoring; Slice Sampling; Bootstrap; Monitoring Intervals; Power Lindley distribution\\\\
{\bf 2020 Mathematical Subject Classification (MSC)}: 62F10, 62F15, 62F40, 62K05, 62N02
 \section{Introduction}
 \indent The variance, and its positive square root -- the standard deviation, have been widely used to study the spread of various datasets about the mean of the data. However, using these measures to compare datasets with different means, distinct measurement units or scales can result in inaccurate conclusions. Therefore, statisticians have over the years developed methods to address these limitations by measuring the variability in proportions relative to the central value of the data. One such type of measure is known as Pearson's coefficient of variation, $C_p$. It is mathematically computed as the ratio of the standard deviation ($\sigma$) relative to the mean ($\mu$), that is, $C_p = {\sigma}/{\mu}$. As a dimensionless quantity, this measure enables meaningful comparisons of the variability of different datasets or populations and is widely regarded as an index of measurement reliability \cite{OS}. Therefore, it has been applied across various fields, and some recent applications are highlighted as follows. In finance, Lei and Zhou \cite{XL} have utilised the $C_p$ of daily trading volume as a volatility measure to study short-term price or volume dynamics. In quality control, Rao et al. \cite{GSR} have developed $C_p$-based control charts for monitoring process variability relative to the mean. In biological sciences, Islam et al. \cite{MZI} have reported phenotypic and genotypic $C_p$'s to quantify trait variability across genotypes. In structural reliability analysis, Szpetnar  \cite{KSS} have utilized $C_p$ estimates for reliability empirical study. Chumnaul \cite{JC} constructed generalized confidence interval for $C_p$ of gamma distribution to measure PM2.5 dispersion. The $C_p$ can also be used to compute the process capability or lifetime performance of products $C_L$ as $C_L = ({1-L/\mu})/{C_p}$, where, $L$ is the lower specification limit and $\mu$ is the lifetime mean. This expression highlights the significant relationship between the performance or capability indices and relative dispersion of a population. In particular, a higher value of the measure $C_p$ corresponds to reduced performance of lifetime products, and vice versa.\\
\indent An alternative measure proposed by Kvålseth \cite{TOK} is known as the second-order coefficient of variation, and is defined as the ratio of the standard deviation relative to the square root of the second moment $\mu'$ about the origin. It can also be expressed in terms of $C_p$ as:
 \begin{equation} \tag{1.1} \label{1.1}
 	C_k = \frac{\sigma}{\sqrt{\mu'}} = \sqrt{\frac{C_p^2}{C_p^2 + 1}}.
 \end{equation}
 $C_k$ is well-defined and its value is restricted to $0 < C_k \leq 1$, with $C_k= 1$ if and only if the mean is zero. This bounded measure can easily interpret data variability and summarize it. In the next paragraph, the lifetime probability model of interest for this article is discussed. \\
\indent The power Lindley distribution, introduced by Ghitany et al. \cite{MEG}, has recently gained attention for its utility as a lifetime distribution. It can be obtained directly from the Lindley distribution as follows. Let $Y \sim \text{Lindley}(\lambda)$, where $\lambda > 0$ is the scale parameter. A random variable $T$ follows the power Lindley distribution if $T=Y^{1/\alpha}$, where $\alpha > 0$ is the shape parameter. Denote the power Lindley (pL) distributed random variable $T$ as $T\sim \text{pL}(\alpha,\lambda)$. The probability density function of $T$ is given as follows:
 \begin{equation}
 	f_T(t) = \begin{cases}
 		\frac{\alpha\lambda^2}{\lambda + 1} t^{\alpha-1} (1 + t^\alpha) e^{-\lambda t^\alpha}, & t > 0 \\
 		0, & \text{otherwise}.
 	\end{cases} \tag{1.2} \label{1.2}
 \end{equation}
 The function $f_T(t)$ can also be represented as a convex mixture of Weibull and generalized gamma density functions.
 Furthermore, the cumulative distribution function of the power Lindley distribution is given as:
 \begin{equation}  \label{1.3} \tag{1.3}
 	F_T(t) = \begin{cases}
 		1 - (1 + \frac{\lambda t^\alpha}{\lambda + 1})e^{-\lambda t^\alpha},& t > 0 \\
 		0, & \text{otherwise.}
 	\end{cases}
 \end{equation}
 The failure rate function is given by:
 \begin{equation}
 	h_T(t) = \begin{cases}
 		\alpha \lambda^2 \frac{(1+t^\alpha)t^{\alpha-1}}{\lambda(1+t^\alpha) + 1},  & t>0 \\
 		0, &  \text{otherwise.}
 	\end{cases}\label{1.4} \tag{1.4}
 \end{equation}
 The measures $C_p$ and $C_k$ for the power Lindley distribution are respectively given as:
 \begin{equation}  \label{1.5} \tag{1.5}
 	C_p = \sqrt{\frac{2\alpha^2 \Gamma(\frac{2}{\alpha})[\alpha(\lambda+1)+2](\lambda + 1)}{\Gamma^2(\frac{1}{\alpha})[\alpha(\lambda + 1) + 1]^2} - 1},~\&~
 	C_k = \sqrt{1 - \frac{\Gamma^2(\frac{1}{\alpha})[\alpha(\lambda + 1) + 1]^2}{2\alpha^2 \Gamma(\frac{2}{\alpha})[\alpha(\lambda+1)+2](\lambda + 1)}}. 
 \end{equation}
 In lifetime studies, making complete observations of failure times can make the experiments sensitive to cost, time, risks, and poor maintenance. Therefore, in such scenarios, it is common for practitioners to collect censored data. Censoring is a practical phenomenon when the exact failure time of a unit is only partially observed. For example, in interval censoring, failures are known to occur only within predetermined monitoring time intervals. When this scheme is followed by controlled withdrawals at the monitoring times, it is known as the progressive type-I interval or simply progressive interval censoring scheme \cite{RA}. This scheme offers better practical flexibility than the interval censoring scheme. Consider an experiment for life testing on $n$ units at prefixed monitoring times $0 = t_0 < t_1 < ... < t_r < t_{r+1} = \infty$. A certain number $\delta_i$ of units fail within each interval $({t_{i-1},~t_i}]$, and at each monitoring time $t_i$, $W_i$ number of units are withdrawn from the experiment. The process of withdrawing the units can be pre-specified by setting the proportions of remaining survived units at each $t_i$ for $i=1,2,...,r$, although withdrawals in most practical scenarios are usually beyond the control of the experimenter. Several authors have studied statistical estimation for certain populations under this censoring scheme, with some recent examples being the inverse Weibull \cite{SS}, inverse Gaussian \cite{SR2}, generalized inverted exponential \cite{RH}, exponential-Poisson \cite{AM} and Weibull \cite{BMN} distributions. Other authors have also focused on optimal monitoring and censoring plans for this censoring scheme \cite{SS, XZ, AK}. \\
 \indent Very few authors have worked on inferential studies for power Lindley distribution using censored data, and these include only the progressive type-II \cite{AK1} and hybrid censoring \cite{KD}. The existing literature does not yet contain any work for estimating the parameters of the power Lindley distribution under the progressive type-I interval censoring scheme. Therefore, this article aims to fill this gap, with primary attention given to estimating $C_p$ and $C_k$. \\
 \indent The various sections of the article are organized as follows. In Section 2, the methods of estimation are discussed, which include the midpoint approximation, the method of moments, the maximum likelihood method, the non-linear least-squares method, and the Bayesian estimation method. In the subsection on maximum likelihood estimation, the method used to obtain the asymptotic confidence intervals based on normal approximation and log-transformation are also discussed. Similarly, in the  Bayesian estimation subsection, the slice sampling algorithm to obtain the credible and highest posterior density intervals are discussed. A discussion on obtaining the optimal monitoring times is given in Section 3. Sections 4 and 5 discuss the data analysis, which respectively include a comprehensive simulation study and an illustration using a real dataset, to compare the resulting performance of the estimators.
 \section{Methods of Estimation}\label{Sec2}
 \indent	Consider a progressive type-I interval censored sample denoted as $\{({t_{i-1},~t_i}], \delta_i, W_i\}_{i=1}^{r}$, from the power Lindley distribution with parameters $(\alpha,\lambda)$. Let $\boldsymbol{t}=(t_1,t_2,...,t_r)$, $\boldsymbol{\delta}=(\delta_1,\delta_2,...,\delta_r)$ and $\boldsymbol{W}=(W_1,W_2,...,W_r)$. Let $\psi$ represent $\alpha$, $\lambda$, $C_p$ or $C_k$.
 \subsection{Mid-Point Approximation} 
 \indent Ng and Wang \cite{HKTN} have analyzed the mid-point estimation of the Weibull parameters by maximizing the log-likelihood function based on a pseudo-right censored sample obtained from a progressive type-I interval censored sample. The failures are assumed to occur at the midpoint $m_i=\frac{t_{i-1}+t_i}{2}$ of the interval $({t_{i-1}, t_i}],~\forall~i=1,2,..., r$. However, the remaining survivors that are right censored at $t_i$ are presumed to fail within the interval $(t_i,~\infty)$. Let $\boldsymbol{m}=(m_1,m_2,...,m_r)$. The pseudo maximum likelihood or mid-point estimators are obtained by directly maximizing the log-likelihood function \eqref{2.1} given as follows:
 \begin{align}
 	\log{\mathscr{L}_m(\alpha,\lambda~|~\boldsymbol{m},\boldsymbol{t},\boldsymbol{\delta},\boldsymbol{W})} \propto~[\log\alpha + 2\log\lambda - \log(\lambda+1)]\sum_{i=1}^{r}\delta_i + \sum_{i=1}^{r} \delta_i\log (1 + m_i^\alpha) + &  \nonumber \\ (\alpha-1)\sum_{i=1}^{r}\delta_i \log m_i +\sum_{i=1}^{r} W_i \log\bigg(1+\frac{\lambda t_i^\alpha}{\lambda + 1}\bigg) - \lambda\sum_{i=1}^{r}(\delta_i m_i^\alpha + W_i t_i^\alpha).& \label{2.1} \tag{2.1}
 \end{align}
 On partial differentiation of Equation \eqref{2.1} with respect to $\alpha$ and $\lambda$, and equating to zero, the following nonlinear equations are obtained:
 \begin{align}
 	\frac{\sum\limits_{i=1}^{r}\delta_i}{\hat{\alpha}^{(m)}} -  {\hat{\lambda}^{(m)}}\sum\limits_{i=1}^{r} (\delta_i m_i^{\hat{\alpha}^{(m)}}\log m_i + W_i t_i^{\hat{\alpha}^{(m)}} \log t_i) + \sum\limits_{i=1}^{r}\frac{\delta_i m_i^{\hat{\alpha}^{(m)}} \log m_i}{1 + m_i^{\hat{\alpha}^{(m)}}} +\sum_{i=1}^{r}\delta_i\log m_i +& \frac{{\hat{\lambda}^{(m)}}}{{\hat{\lambda}^{(m)}} + 1}\nonumber \\
 	\times \sum_{i=1}^{r} \frac{W_i t_i^{\hat{\alpha}^{(m)}} \log t_i}{1 + \frac{{\hat{\lambda}^{(m)}} t_i^{\hat{\alpha}^{(m)}}}{{\hat{\lambda}^{(m)}} + 1}} = 0,&\label{2.2}\tag{2.2}\\
 	{(\hat{\lambda}^{(m)} + 2)\sum\limits_{i=1}^{r}\delta_i} - {\hat{\lambda}^{(m)}(\hat{\lambda}^{(m)} + 1)}\sum_{i=1}^{r} (\delta_i m_i^{\hat{\alpha}^{(m)}} + W_i t_i^{\hat{\alpha}^{(m)}} ) + \sum_{i=1}^{r} \frac{W_i t_i^{\hat{\alpha}^{(m)}}}{(1 + t_i^{\hat{\alpha}^{(m)}}) + \frac{1}{\hat{\lambda}^{(m)}}} = 0.& \label{2.3}\tag{2.3}
 \end{align}
 where, $\hat{\alpha}^{(m)}$ and $\hat{\lambda}^{(m)}$ denote the mid-point estimators of $\alpha$ and $\lambda$ respectively. It is readily observed that $\hat{\alpha}^{(m)}$ and $\hat{\lambda}^{(m)}$ cannot be obtained analytically. Therefore, these non-linear equations are solved by utilizing existing root-solving techniques such as the Newton-Raphson method. Various modern numerical computing software programs that are equipped with in-built root-solving packages are easily implemented. By invariance property, the mid-point estimators of $C_p$ and $C_k$ are respectively given by $\hat{C}_p^{(m)}$ and $\hat{C}_k^{(m)}$.
 \subsection{Method of Moments} 
 \indent The method of moments requires that the population moments be equated to their sample counterparts. Let $T\sim\text{pL}(\alpha,\lambda)$. The $k^\text{th}$ moment of $T$ about the origin is given by
 \begin{equation}
 	\nonumber
 	\mathbb{E}(T^k) = \frac{k\Gamma(\frac{k}{\alpha})[\alpha(\lambda+1)+k]}{\alpha^2 \lambda^{\frac{k}{\alpha}}(\lambda + 1) }.  \label{2.4}\tag{2.4}
 \end{equation}
 However, since the given data is progressive type-I interval censored, the $k^\text{th}$ population moments is directly equated to the $k^\text{th}$ moments of a doubly truncated power Lindley distribution for an interval $(T_l,~T_u]$ defined as:
 \begin{equation} \label{2.5}\tag{2.5}
 	\mathbb{E}[T^k~|~T\in {(T_l,~T_u]}] = \frac{\int\limits_{T_l}^{T_u} t^k f(t)}{F(T_u) - F(T_l)}.
 \end{equation}
 The estimators of $\alpha$ and $\lambda$ by the method of moments are denoted as $\check{\alpha}$ and $\check{\lambda}$ respectively, and are obtained by solving the following system of non-linear equations:
 \begin{align}
 	\label{2.6}\tag{2.6}
 	\frac{\Gamma(\frac{1}{\check{\alpha}})[{\check{\alpha}}({\check{\lambda}}+1)+1]}{{\check{\alpha}}^2 {\check{\lambda}}^{\frac{1}{{\check{\alpha}}}}({\check{\lambda}} + 1) } &= \frac{1}{n}\sum_{i=1}^{r}\{\delta_i \mathbb{E}[T~|~T\in {(t_{i-1},~t_i]}] + W_i \mathbb{E}[T~|~T\in {(t_i,~\infty)}]\}, \\
 	\label{2.7}\tag{2.7}
 	\frac{2\Gamma(\frac{2}{{\check{\alpha}}})[{\check{\alpha}}({\check{\lambda}}+1)+2]}{{\check{\alpha}}^2 {\check{\lambda}}^{\frac{2}{{\check{\alpha}}}}({\check{\lambda}} + 1) } &= \frac{1}{n}\sum_{i=1}^{r}\{\delta_i \mathbb{E}[T^2~|~T\in {(t_{i-1},~t_i]}] + W_i \mathbb{E}[T^2~|~T\in {(t_i,~\infty)}] \}. 
 \end{align} 
 Since the moment estimates cannot be solved in closed form, therefore, the root-solving techniques to obtain $(\check{\alpha},\check{\lambda})$ are used. Naturally, the moment estimators $\check{\alpha}$ and $\check{\lambda}$ can be plugged into the Equation \eqref{1.5} to obtain the moment estimators $\check{C}_p$ and $\check{C}_k$ respectively. However, since the censoring scheme forces the sample moments to be replaced by their expected conditional sample moments, the resulting estimators may have finite sample bias.
 \subsection{Maximum Likelihood Estimation} 
 \indent In the frequentist method of maximum likelihood estimation, the resulting maximum likelihood estimators (MLEs) of $\alpha$ and $\lambda$ are obtained by maximizing the likelihood function given as:
 \begin{equation}
 	\mathscr{L} (\alpha,\lambda~|~\boldsymbol{t},\boldsymbol{\delta},\boldsymbol{W}) \propto \prod_{i=1}^{r} [F_T(t_i~|~\alpha,\lambda) - F_T(t_{i-1}~|~\alpha,\lambda)]^{\delta_i} [1 - F_T(t_i~|~\alpha,\lambda)]^{W_i}. \label{2.8}\tag{2.8}
 \end{equation}
 Taking natural logarithm of Equation \eqref{2.8}, the log-likelihood function obtained is reduces to:
 \begin{equation} \label{2.9}\tag{2.9}
 	\log\mathscr{L} (\alpha,\lambda~|~\boldsymbol{t},\boldsymbol{\delta},\boldsymbol{W}) \propto \sum_{i=1}^{r}\delta_i\log D_i + \sum_{i=1}^{r}W_i \log\bigg(1 + \frac{\lambda t_i^\alpha}{\lambda + 1}\bigg) - \lambda \sum_{i=1}^{r} W_i t_i^\alpha,
 \end{equation}
 where, $D_i = \bigg(1 + \frac{\lambda t_{i-1}^\alpha}{\lambda + 1}\bigg)e^{-\lambda t_{i-1}^\alpha} - \bigg(1 + \frac{\lambda t_i^\alpha}{\lambda + 1}\bigg)e^{-\lambda t_i^\alpha}$, $i=1,2,...,r$. \\
 \indent The approaches to maximizing Equation \eqref{2.9} are discussed as follows:\\\\
 \textbf{A. Newton-Raphson Method} \\\\
 Differentiating Equation \eqref{2.9} with respect to $\alpha$ and $\lambda$, and further equating to 0 the log-likelihood Equations \eqref{2.10} and \eqref{2.11} are obtained. These equations can be solved numerically to obtain the maximum likelihood estimators $\hat{\alpha}$ and $\hat{\lambda}$:
 \begin{align}
 	\frac{\partial}{\partial\alpha}\log\mathscr{L}(\alpha,\lambda~|~\boldsymbol{t},\boldsymbol{\delta},\boldsymbol{W}) &= \sum_{i=1}^{r}\delta_i \frac{ D_{i,\alpha}}{D_i} + \frac{\lambda}{\lambda + 1}\sum_{i=1}^{r}W_i \frac{t_i^\alpha\log t_i}{\big(1 + \frac{\lambda t_i^\alpha}{\lambda + 1} \big)} - \lambda \sum_{i=1}^{r}W_i t_i^\alpha \log t_i = 0, \label{2.10} \tag{2.10} \\
 	\frac{\partial}{\partial\lambda} \log\mathscr{L}(\alpha,\lambda~|~\boldsymbol{t},\boldsymbol{\delta},\boldsymbol{W})&= \sum_{i=1}^{r} \delta_i \frac{D_{i,\lambda}}{D_i} + \frac{1}{(\lambda+1)^2}\sum_{i=1}^{r}W_i\frac{t_i^\alpha}{\big(1 + \frac{\lambda t_i^\alpha}{\lambda + 1} \big)}  - \sum_{i=1}^{r}W_i t_i^\alpha= 0,\label{2.11} \tag{2.11} 
 \end{align}
 where $D_{i,\alpha} = \frac{\partial D_i}{\partial \alpha}$ and $D_{i,\lambda} = \frac{\partial D_i}{\partial \lambda}$. These two non-linear equations cannot be solved analytically. Therefore, numerical root finders can be used, such as the Newton-Raphson's algorithm. However, this simultaneous approach for obtaining the two parameter estimates is sensitive to the initial values, and in some cases can bring about a slow convergence to the solution. Many alternative algorithms such as the EM algorithm, have slow convergence rates, and sometimes even they can fail to converge to the appropriate values. \\\\
 \textbf{B. Proposed Transformation Approach} \\\\
 Here, an alternative approach for the MLEs is proposed by simply transforming the two-parameter optimization problem to a single-parameter one. It is clear by definition that if $T\sim\text{pL}(\alpha,\lambda)$, then $Y = T^\alpha\sim\text{Lindley}(\lambda)$. If $\alpha$ is known, one can easily reduce the estimation problem to one for a Lindley distribution under progressive type-I interval censoring. That is, the monitoring times $\{({t_{i-1},~t_i}], \delta_i, W_i\}_{i=1}^{r}$ are transformed to $\{({y_{i-1},~y_i}], \delta_i, W_i\}_{i=1}^{r}$, where $y_i = t_i^\alpha$. The estimator for $\lambda$ is then obtained by replacing the midpoints of Equation \eqref{2.3} by the conditional expected failure time $y_{({y_{i-1}, y_i}]}' = \mathbb{E}[Y ~|~Y\in {(y_{i-1},~y_i]}]$ and solve the resulting equation
 \begin{equation}
 	{(\hat{\lambda} + 2)\sum\limits_{i=1}^{r}\delta_i} - {\hat{\lambda}(\hat{\lambda} + 1)}\sum_{i=1}^{r} (\delta_i y_{({y_{i-1}, y_i}]}' + W_i y_i ) + \sum_{i=1}^{r} \frac{W_i y_i}{(1 + y_i) + \frac{1}{\hat{\lambda}}} = 0,\label{2.12}\tag{2.12}
 \end{equation}
 where,
 \begin{equation} \label{2.13}\tag{2.13}
 	y_{({y_{i-1},~y_i}]}' =   \frac{\{\lambda^2 y_{i-1}(1+y_{i-1}) + \lambda(1+2y_{i-1})+2\}e^{-\lambda y_{i-1}} - \{\lambda^2 y_i(1+y_i) + \lambda(1+2y_i)+2\}e^{-\lambda y_i}}{{\lambda}[\{\lambda(1+y_{i-1})+1\}e^{-\lambda y_{i-1}} - \{\lambda(1+y_{i})+1\}e^{-\lambda y_{i}} ]}.
 \end{equation}
 \begin{algorithm}[b!]
 	\caption{(Transformation-based Maximum Likelihood Estimation)}\label{Al2}
 	\begin{algorithmic}[1]
 		\State Fix $\varepsilon_\alpha$ for $\alpha$ and $\varepsilon_\lambda$ for $\lambda$ and initialize $\alpha_0=\hat{\alpha}^{(m)}$.
 		\While {$|\hat{\alpha}-\alpha_0| \geq \varepsilon_\alpha$} \textbf{do}
 		\State Transform $y_i = t_i^\alpha$ at $\alpha=\alpha_0$ for $i=1,2,...,r$. 
 		\State  Initialize $\lambda_0$.
 		\While {$|\hat{\lambda}-\lambda_0| \geq \varepsilon_\lambda$} \textbf{do} 
 		\State Calculate $y_{({y_{i-1}, y_i}]}'$. 
 		\State Solve for $\hat{\lambda}$ from Equation \eqref{2.12}. 
 		\If {$|\hat{\lambda}-\lambda_0| < \varepsilon_\lambda$} 
 		\State Exit.
 		\Else
 		\State $\lambda_0 = \hat{\lambda}$.
 		\EndIf
 		\EndWhile
 		\State Compute $\hat{\alpha} = \arg \max\limits_{\alpha} \log \mathscr{L}(\hat{\lambda},\alpha)$.
 		\If{$ |\hat{\alpha}-\alpha_0| < \varepsilon_\alpha$}
 		\State Exit. 
 		 \Else
 		\State $\alpha_0 = \hat{\alpha}$.
 		\EndIf
 		\EndWhile
 	\end{algorithmic}
 \end{algorithm}
 The obtained $\hat{\lambda}$ is then substituted in the log-likelihood function \eqref{2.9}, which is then maximized to obtain $\hat{\alpha}$. This process is repeated until the log-likelihood function is maximized up to a given tolerance value for each estimate. The proper flow of this method is detailed in Algorithm \ref{Al2}. The initialization of $\lambda$ can be achieved by solving the mid-point Equation \eqref{2.3}, with $m_i = (y_{i-1}+y_i)/2$. By the invariance property, the MLEs of $C_p$ and $C_k$ are respectively given and denoted by $\hat{C}_p$ and $\hat{C}_k$. \\
 \indent When the sample size is very large, the MLEs asymptotically follow normal distributions. That is, $\hat{\alpha} \sim N(\alpha,\hat{\sigma}_{\hat{\alpha}}^2)$ and $\hat{\lambda} \sim N(\lambda,\hat{\sigma}_{\hat{\lambda}}^2)$. The asymptotic variances for these parameters are obtained through the covariance matrix $\hat{V}$ which is the inverse of the expected Fisher's information matrix $\hat{\boldsymbol{I}}$:
 \begin{equation} \nonumber
 	\hat{V} = \begin{bmatrix}
 		\hat{\sigma}_{\alpha}^2 & \widehat{\text{cov}}(\alpha,\lambda) \\
 		\widehat{\text{cov}}(\lambda,\alpha) & \hat{\sigma}_{\lambda}^2
 	\end{bmatrix}_{(\alpha = \hat{\alpha},~\lambda=\hat{\lambda})} = \hat{\boldsymbol{I}}^{-1}= \begin{bmatrix}
 		\mathbb{E}\{-\frac{\partial^2\log\mathscr{L}(\alpha,\lambda)}{\partial \alpha^2}\} & \mathbb{E}\{-\frac{\partial^2\log\mathscr{L}(\alpha,\lambda)}{\partial \alpha\partial\lambda}\} \\
 		\mathbb{E}\{-\frac{\partial^2\log\mathscr{L}(\alpha,\lambda)}{\partial \lambda\partial\alpha}\} & \mathbb{E}\{-\frac{\partial^2\log\mathscr{L}(\alpha,\lambda)}{\partial \lambda^2}\}
 	\end{bmatrix}_{(\alpha = \hat{\alpha},~\lambda=\hat{\lambda})}^{-1}. \label{2.13}\tag{2.13}
 \end{equation}
 Furthermore, to obtain the 100$(1-\eta)$\% standard asymptotic confidence intervals for the coefficients of variation, the Delta method \cite{WQM} is employed to acquire the variances of their MLEs: $\hat{\sigma}_{\hat{C_p}}^2 = \{[\nabla {C_p}]^T [\hat{V}] [\nabla {C_p}]\}_{(\hat{\alpha},\hat{\lambda})}$ and $\hat{\sigma}_{\hat{C_k}}^2 = \{[\nabla {C_k}]^T [\hat{V}] [\nabla {C_k}]\}_{(\hat{\alpha},\hat{\lambda})}$.
 Therefore, the 100$(1-\eta)$\% standard asymptotic confidence intervals for the parameter $\psi$, where $\psi = \alpha,\lambda, C_p$ and $C_k$, is given by $(\hat{\psi} \mp Z_{1-\frac{\eta}{2}}\sqrt{\hat{\sigma}_{\hat{\psi}}^2})$, where $Z_{1-\frac{\eta}{2}}$ is the upper $100(1-\frac{\eta}{2})^\text{th}$ percentile of the standard normal distribution. The lower bounds of these intervals can in fact be negative in some cases, and therefore inappropriate for the parameters which are strictly positive. Therefore, the corrected asymptotic confidence intervals ($I_\text{AC}$) for the parameter $\psi$ is given by:
 \begin{equation}
 	\nonumber
 	[\max(0,\hat{\psi} - Z_{1-\frac{\eta}{2}}\sqrt{\hat{\sigma}_{\hat{\psi}}^2}),\hat{\psi} + Z_{1-\frac{\eta}{2}}\sqrt{\hat{\sigma}_{\hat{\psi}}^2}].
 \end{equation} 
 Alternatively, the MLE $\hat{\psi}$ is log-transformed as:
 \begin{equation}
 	\nonumber
 	\frac{\log \hat{\psi}-\log \psi}{{{\hat{\sigma}_{\log \hat{\psi}}}}} \sim N(0,1).
 \end{equation} 
 Here, $\hat{\sigma}_{\log{\hat{\psi}}}^2 = \frac{\hat{\sigma}_{{\hat{\psi}}}^2}{\hat{\psi}^2}$, obtained using the Delta method for $\log \psi$. Therefore, a $100(1-\eta)\%$ asymptotic confidence interval ($I_\text{LAC}$) for $\psi$ based on log-transformation of the MLEs, is obtained as follows:
 \begin{equation} \nonumber
 	\bigg[\hat{\psi} \exp\bigg\{-\frac{Z_{1-\frac{\eta}{2}} \sqrt{\hat{\sigma}^2_{\hat{\psi}}}}{\hat{\psi}}\bigg\},~\hat{\psi} \exp\bigg\{\frac{Z_{1-\frac{\eta}{2}} \sqrt{\hat{\sigma}^2_{\hat{\psi}}}}{\hat{\psi}}\bigg\}\bigg].
 \end{equation}
 	\subsection{Non-Linear Least Squares Estimation} 
 The direct linear least squares or probability plot estimation for the power Lindley distribution is not straightforward, due to the sophisticated structure of the cdf $F_T(t)$. Therefore, an alternative non-linear type of least squares estimation method is proposed whereby the sum of squared-errors between the theoretical and the observed cdfs are minimized. The corresponding failures and withdrawals are adjusted as weights in the objective function defined as follows:
 \begin{align}
 	\min_{(\alpha,\lambda)}	\Phi_0 &= \sum_{i=1}^{r} \delta_i[\{F_T(t_i~|~\alpha,\lambda)-F_T(t_{i-1}~|~\alpha,\lambda)\} - \{\bar{F}_i - \bar{F}_{i-1}\}]^2 + \sum_{i=1}^{r} W_i[F_T(t_i~|~\alpha,\lambda) - \bar{F}_i]^2,  \nonumber\\
 	&= \sum_{i=1}^{r}\delta_i [D_i - (\bar{F}_i - \bar{F}_{i-1})]^2 + \sum_{i=1}^{r}W_i\bigg[1-\bigg(1+\frac{\lambda}{\lambda+1} t_i^\alpha\bigg)e^{-\lambda t_i^\alpha} -  \bar{F}_i\bigg]^2. \nonumber \\
 	&\text{subject to}~\alpha>0, \lambda>0. \tag{2.14}\label{2.14}
 \end{align}
 where, $\bar{F}_i$ is the non-parametric estimator of $F_T(t_i)$, and it can be obtained in various ways, for example, the product-limit and the moments-approximation method \cite{HHQ}. In this subsection, the latter is employed:
 \begin{equation}   \tag{2.15}\label{2.15}
 	\bar{F}_i = 1 - \prod_{k=r-i+1}^{r} \bigg[\frac{\sum_{j=r-k+2}^{r}\delta_j + \sum_{j=r-k+1}^{r}W_j + i}{{\sum_{j=r-k+1}^{r}(\delta_j + W_j) + i+1}} \bigg]~~\forall~i=1,2,...,r.
 \end{equation}
 To minimize $\Phi_0$, the following system of non-linear equations are solved for the non-linear least squares estimators $(\tilde{\alpha},\tilde{\lambda})$ of the parameters.
 \begin{align}
 	\sum_{i=1}^{r}\delta_i[\tilde{D}_i - (\bar{F}_i-\bar{F}_{i-1})] \tilde{D}_{i,\alpha} &= \frac{\tilde{\lambda}}{\tilde{\lambda}+1}\sum_{i=1}^{r} W_i(t_i^{\tilde{\alpha}}+t_i^{2\tilde{\alpha}})\log t_i e^{-\tilde{\lambda}t_i^{\tilde{\alpha}}}\bigg[(1+\frac{\tilde{\lambda}}{\tilde{\lambda}+1} t_i^{\tilde{\alpha}})e^{-\tilde{\lambda} t_i^{\tilde{\alpha}}} - \bar{S}_i\bigg], \nonumber\\
 	\sum_{i=1}^{r}\delta_i[\tilde{D}_i - (\bar{F}_i-\bar{F}_{i-1})] \tilde{D}_{i,\lambda} &=\sum_{i=1}^{r}  W_i\frac{t_i^{\tilde{\alpha}} e^{-\tilde{\lambda} t_i^{\tilde{\alpha}}}}{(\tilde{\lambda}+1)^2}[(\tilde{\lambda}+1)(\tilde{\lambda}+1-\tilde{\lambda} t_i^{\tilde{\alpha}})-1]\bigg[(1+\frac{\tilde{\lambda}}{\tilde{\lambda}+1} t_i^{\tilde{\alpha}})e^{-\tilde{\lambda} t_i^{\tilde{\alpha}}} - \bar{S}_i\bigg]. \nonumber
 \end{align}
 This optimization problem can numerically be unstable, as the two parameters are constrained to $\mathbb{R}^+$. Therefore, such unwanted issues can be fixed by log-transformation $\log \alpha = {\xi_\alpha}$ and $\log\lambda = {\xi_\rho}$, so that the unconstrained minimization problem becomes:
 \begin{equation} \label{2.16}\tag{2.16}
 	\min_{(\alpha,\lambda)} \Phi_\xi = \sum_{i=1}^{r}\delta_i [D_{i,\xi} - (\bar{F}_i - \bar{F}_{i-1})]^2 + \sum_{i=1}^{r}W_i\bigg[1-\bigg(1+\frac{e^{\xi_\lambda}}{e^{\xi_\lambda}+1} t_i^{e^{\xi_\alpha}}\bigg)e^{-e^{\xi_\lambda} t_i^{e^{\xi_\alpha}}} -  \bar{F}_i\bigg]^2 ,
 \end{equation}
 where, $D_{i,\xi} = \bigg(1+\frac{e^{\xi_\lambda} t_{i-1}^{e^{\xi_\alpha}}}{e^{\xi_\alpha} + 1}\bigg)e^{-e^{\xi_\lambda} t_{i-1}^{e^{\xi_\alpha}}} - \bigg(1+\frac{e^{\xi_\lambda} t_i^{e^{\xi_\alpha}}}{e^{\xi_\alpha} + 1}\bigg)e^{-e^{\xi_\lambda} t_i^{e^{\xi_\alpha}}}$. On plugging-in $\tilde{\alpha}$ and $\tilde{\lambda}$, the non-linear least squares estimates of $C_p$ and $C_k$ are respectively obtained as $\tilde{C}_p$ and $\tilde{C}_k$.
  \begin{algorithm}[b!]
 	\caption{(Percentile Bootstrap)}\label{Al3}
 	\begin{algorithmic}[1]
 		\State Obtain the MLE $\hat{\alpha}$ and $\hat{\lambda}$ using the given $\{({t_{i-1}, t_i}], \delta_i, W_i\}_{i=1}^{r}$. 
 		\State Set $S_\text{boot}$ and let $s=1$.
 		\For{$s=1,2,...,S_\text{boot}$}
 		\State Under the same censoring plan, generate $\{({t_{i-1}, t_i}], {\delta}_i^*, {W}_i^*\}_{i=1}^{r}$ using $(\hat{\alpha},\hat{\lambda})$.
 		\State Determine the bootstrap likelihood estimates $\hat{\psi}_s$, where $\psi =\alpha,\lambda,C_p$ and $C_k$.
 		\EndFor
 		\State Sort $\hat{\psi}_{(1)} \leq \hat{\psi}_{(2)} \leq ...\leq \hat{\psi}_{(S_\text{boot})}$.
 		\State The bootstrap point estimate is $\hat{\psi}^{(b)} = \frac{1}{S_\text{boot}}\sum\limits_{i=1}^{S_\text{boot}} \hat{\psi}_{(s)}.$
 		\State A $100(1-\eta)\%$ ${I}_{\text{PB}}$ is $\big(\psi_{(\frac{\eta S_\text{boot}}{2})}^{(b)}, \psi_{(\{1 - \frac{\eta}{2}\} S_\text{boot} )}^{(b)}\big)$.
 	\end{algorithmic}
 \end{algorithm}
 \subsection{Bootstrap Estimation}
 Introduced by Efron \cite{BE}, bootstrapping is a repeated resampling procedure with replacement from the original sample. As seen earlier, the MLEs are asymptotically consistent for large sample sizes. The sampling distribution of the MLE can be highly skewed or biased when sample sizes are small, making normal approximations unreliable. This technical drawback affects asymptotic confidence intervals ($I_\text{AC}$), leading to inaccuracies and potentially misleading coverage. Bootstrapping addresses these issues by empirically estimating the estimator's sampling distribution directly from data, without relying on asymptotic assumptions. Even when MLEs are used as the bootstrapping statistics, resampling typically yields more accurate measures of variability and better confidence intervals. In this subsection, the bootstrap point estimate are obtained using the percentile bootstrap method. Two types of bootstrap intervals are also obtained, the percentile bootstrap (${I}_{\text{PB}}$) and the Studentized-t (${I}_{\text{StB}}$) bootstrap intervals. \\ 
 \indent The most basic bootstrapping method is the percentile bootstrap. The bootstrap samples are generated for the power Lindley distribution using the MLEs obtained $(\hat{\alpha},\hat{\lambda})$, and the bootstrap estimate $\hat{\psi}_s$ of $\psi =\alpha,\lambda,C_p$ and $C_k$, is obtained correspondingly to each bootstrap sample number $s$. The Algorithm \ref{Al3} provides the flow to obtain the bootstrap estimate and ${I}_{\text{PB}}$. \\
 \indent An alternative to the percentile bootstrap is the Student t-bootstrap. This method provides more exact and less biased confidence intervals compared to the ${I}_{\text{PB}}$. The steps of this method are similar to the percentile bootstrap method, but the t-statistic $\dot{t}_s = \frac{\hat{\psi}_s - \hat{\psi}}{\sqrt{\hat{\sigma}_{\hat{\psi}_s}^2}}$ is calculated for each bootstrap estimate. Here, the bootstrap standard error estimate $\hat{\sigma}_{\hat{\psi}_s}^2$ is obtained using the Delta method. The flow of this method is given in Algorithm \ref{Al4}.
 \begin{algorithm}[t!]
 	\caption{(Student t-Bootstrap)}\label{Al4}
 	\begin{algorithmic}[1]
 		\State Obtain the MLE $\hat{\alpha}$ and $\hat{\lambda}$ using the given $\{({t_{i-1}, t_i}], \delta_i, W_i\}_{i=1}^{r}$. 
 		\State Set $S_\text{boot}$ and let $s=1$.
 		\For{$s=1,2,...,S_\text{boot}$}
 		\State Under the same censoring plan, generate $\{({t_{i-1}, t_i}], \delta_i^*, W_i^*\}_{i=1}^{r}$ using $(\hat{\alpha},\hat{\lambda})$.
 		\State Determine $\hat{\psi}_s$, for $\psi =\alpha,\lambda,C_p$ and $C_k$.
 		\State Determine $\hat{\sigma}_{\hat{\psi}_s}^2$ using Delta method.
 		\State Compute $\dot{t}_s = \frac{\hat{\psi}_s - \hat{\psi}}{\sqrt{\hat{\sigma}_{\hat{\psi}_s}^2}}$.
 		\EndFor
 		\State Sort $\dot{t}_{(1)} \leq \dot{t}_{(2)} \leq ...\leq \dot{t}_{(S_\text{boot})}$.
 		\State A $100(1-\eta)\%$ ${I}_{\text{StB}}$ is $\big(\hat{\psi} - \dot{t}_{(\{1 - \frac{\eta}{2}\} S_\text{boot} )}\sqrt{\hat{\sigma}_{\hat{\psi}}^2}, \hat{\psi} - \dot{t}_{(\frac{\eta S_\text{boot}}{2})}\sqrt{\hat{\sigma}_{\hat{\psi}}^2}\big)$.
 	\end{algorithmic}
 \end{algorithm}
 \subsection{Bayesian Estimation}
 Unlike frequentist methods, Bayesian estimation treats unknown parameters as random quantities. These parameters are specified by prior knowledge through a probability distribution. The observed data provides information about the parameters via the likelihood function. This updates the prior to the posterior distribution using Bayes' theorem. For the power Lindley distribution, the parameters $\alpha$ and $\lambda$ have positive real support. The gamma distribution can be independently set as their priors. This leads to the joint prior density:
 \begin{equation} \label{2.17}\tag{2.17}
 	\pi(\alpha,\lambda) \propto \alpha^{a_1-1} \lambda^{a_2-1} \exp(-b_1\alpha-b_2\lambda),
 \end{equation}
 where $a_i > 0$ and $b_i > 0$ $\forall$ $i=1,2$ denote the prior hyperparameters. In instances when the prior information is not readily available, the hyperparameters can be selected to be 0, but this results in the improper prior known as the Jeffreys' non-informative prior:
 \begin{equation} \label{2.18}\tag{2.18}
 	\pi_{J}(\alpha,\lambda) \propto \frac{1}{\alpha\lambda}.
 \end{equation}
 Consequently, the joint posterior density function is obtained:
 \begin{equation} \label{2.19}\tag{2.19}
 	\pi_p(\alpha,\lambda~|~\boldsymbol{t},\boldsymbol{\delta},\boldsymbol{W}) = \frac{\mathscr{L}(\alpha,\lambda~|~\boldsymbol{t},\boldsymbol{\delta},\boldsymbol{W})\pi(\alpha,\lambda)}{\int\limits_{0}^{\infty}\int\limits_{0}^{\infty} \mathscr{L}(\alpha,\lambda~|~\boldsymbol{t},\boldsymbol{\delta},\boldsymbol{W})\pi(\alpha,\lambda)d\alpha d\lambda},
 \end{equation}
  \begin{align} 
 	\text{where,}\quad\mathscr{L}(\alpha,\lambda~|~\boldsymbol{t},\boldsymbol{\delta},\boldsymbol{W})\pi(\alpha,\lambda) \propto 
 	\exp\bigg[\sum_{i=1}^{r}\delta_i\log D_i + \sum_{i=1}^{r}W_i \log\bigg(1 + \frac{\lambda t_i^\alpha}{\lambda + 1}\bigg) - \lambda \sum_{i=1}^{r} W_i t_i^\alpha + & \nonumber\\
 	(a_1 - 1)\log\alpha + (a_2-1)\log\lambda - b_1\alpha-b_2\lambda\bigg].& \nonumber
 \end{align}
 The Bayes estimator $\hat{\psi}(\alpha,\lambda)$ of any smooth function $\psi(\alpha,\lambda)$ is obtained by estimating the value for which the posterior expected loss is minimum. The form of such an estimator  depends solely on the chosen loss function. In this article, the squared error loss function is considered:
 \begin{equation} \nonumber
 	L_s\{\psi(\alpha,\lambda),\hat{\psi}_s(\alpha,\lambda)\} = [\psi(\alpha,\lambda) - \hat{\psi}_s(\alpha,\lambda)]^2 .
 \end{equation}
 Then, the Bayes' estimator $\hat{\psi}_s(\alpha,\lambda)$ is given by the conditional posterior expectation:
 \begin{equation} 
 	\hat{\psi}_s(\alpha,\lambda) = \int\limits_{0}^{\infty}\int\limits_{0}^{\infty} \psi(\alpha,\lambda){\pi_p}(\alpha,\lambda~|~\boldsymbol{t},\boldsymbol{\delta},\boldsymbol{W}) d\alpha d\lambda = \frac{\int\limits_{0}^{\infty}\int\limits_{0}^{\infty} \psi(\alpha,\lambda) \mathscr{L}(\alpha,\lambda~|~\boldsymbol{t},\boldsymbol{\delta},\boldsymbol{W})\pi(\alpha,\lambda) d\alpha d\lambda }{\int\limits_{0}^{\infty}\int\limits_{0}^{\infty} \mathscr{L}(\alpha,\lambda~|~\boldsymbol{t},\boldsymbol{\delta},\boldsymbol{W})\pi(\alpha,\lambda)d\alpha d\lambda}. \tag{2.20} \label{2.20}
 \end{equation}
  \begin{algorithm}[b!]
 	\caption{(Slice Sampling Algorithm)}\label{Al5}
 	\begin{algorithmic}[1]
 		\State Initialize $\alpha^{(0)}$ and $\lambda^{(0)}$.
 		\State Fix $H$, the burn-in $H_b$, and let $h=1$.
 		\While{$h \leq H$} 
 		\State Keep $\lambda^{(h-1)}$ fixed, draw $z_\alpha~\sim~\mathscr{U}(\mathscr{R}_h(\alpha))$ and let $y_\alpha = \log z_\alpha$.
 		\State Determine $\mathscr{I}_\alpha$ by conditioning on $y_\alpha$.
 		\State Generate $\alpha^{(h)} \sim \mathscr{U}(\mathscr{I}_\alpha \cap \mathscr{R}_h (\alpha))$. 
 		\State Keep $\alpha^{(h)}$ fixed, draw $z_\lambda~\sim~\mathscr{U}[\mathscr{R}_h(\lambda)]$ and let $y_\lambda = \log z_\lambda$.
 		\State Determine $\mathscr{I}_\lambda$ by conditioning on $y_\lambda$. 
 		\State Generate $\lambda^{(h)}\sim \mathscr{U}(\mathscr{I}_\lambda\cap \mathscr{R}_h(\lambda))$.
 		\State Calculate $\psi^{(h)}$.
 		\State Set $h=h+1$.
 		\EndWhile
 		\State Determine $\hat{\psi}^{(g)}$ (if using gamma prior) and $\hat{\psi}^{(J)}$ (if using Jeffreys' prior) under SEL function.
 		\State Determine ${I}_{\text{EBC}}(g)$ and ${I}_{\text{HPD}}(g)$ (gamma prior); and ${I}_{\text{EBC}}(J)$ and ${I}_{\text{HPD}}(J)$ (Jeffreys' prior).
 	\end{algorithmic}
 \end{algorithm}
 The ratio of the two integrals on the right does not give out closed-form expressions for $\hat{\psi}(\alpha,\lambda)$. There are several ways to counter this problem, either by numerical approximation of the integrals or by sampling from the posterior distributions through the Monte Carlo simulation approach. In this particular study, the latter is adopted, so that the randomly sampled states are collected from the parameter space, creating a Markov Chain of length $H$, whose stationary distribution converges to the posterior distribution. After a certain burn-in period $H_b$, the Bayes estimator $\hat{\psi}(\alpha,\lambda)$ is the arithmetic mean of the remaining $H-H_b$ samples. There exist many Markov Chain Monte Carlo (MCMC) simulation algorithms, and one such type known as the two-step, single-variable slice sampling approach \cite{RMN} is specifically implemented. Let at any iteration $h$, fix $\lambda^{(h-1)}$, and the first step requires that an auxiliary variable $z_\alpha$ is drawn uniformly over the region $\mathscr{R}_h (\alpha) = \{\alpha : z_\alpha < {\pi_p}(\alpha^{(h-1)},\lambda^{(h-1)})\}$ containing $\alpha^{(h-1)} $. A random interval $\mathscr{I}_\alpha$ is placed around $\alpha^{(h-1)}$ in such a way that it contains $\mathscr{R}_h (\alpha)$ as much as possible. The new state $\alpha^{(h)}$ is then drawn from the new set $\mathscr{I}_\alpha \cap \mathscr{R}_h (\alpha)$. Consequently, fixing $\alpha^{(h)}$, the new sample $\lambda^{(h)}$ is drawn in a similar manner. That is, at the same iteration $h$, fix $\alpha^{(h)}$, draw the auxiliary variable $z_\lambda$ uniformly over the region $\mathscr{R}_h(\lambda) = \{\lambda:z_\lambda < \pi_p(\alpha^{(h)},\lambda^{(h-1)})\}$. Then the random interval $\mathscr{I}_\lambda$ is determined, and the new sample $\lambda^{(h)}$ is drawn uniformly from the set $\mathscr{I}_\lambda \cap \mathscr{R}_h(\lambda)$. The flow of the slice sampling is given in brief in Algorithm \ref{Al5}.
 \\
 \indent Using Bayesian analysis, one can also easily construct the equal-tailed Bayesian credible (${I}_{\text{EBC}}$) and highest posterior density intervals (${I}_{\text{HPD}}$) when the posterior distribution, or the Markov chain whose stationary distribution representing one is readily available. The $100(1-\eta)\%$ ${I}_{\text{EBC}}$ of $\psi =\alpha,\lambda,C_p$ and $C_k$ is given by $[{\psi}_{(\frac{H\eta}{2})},{\psi}_{(H(1-\frac{\eta}{2}))}]$, where ${(\frac{H\eta}{2})}$ denotes the lower $(\frac{H\eta}{2})^\text{th}$ percentile of the posterior distribution of the $\psi$. The ${I}_{\text{HPD}}$ are derived by applying Corollary 1 of Chen and Shao \cite{MHC}. That is, the $100(1-\eta)\%$ ${I}_{\text{HPD}}$ of ${\psi}$ is  $(\psi_{(H')}, \psi_{(H'+H\{1-\eta\})})$, where $H'$ is given by:
 \begin{equation} \tag{2.21} \label{2.21}
 	H' = \arg\min_{1\leq h \leq H\eta} [\psi_{(h + H\{1-\eta\})}- \psi_{(h)}].
 \end{equation}
 \section{Optimal Monitoring Plans} \label{Sec3}
 \indent The choice of monitoring times and the proportions of withdrawals reflect on the overall results of the estimation problems. As discussed earlier, it is often the case that the withdrawals are beyond the control of the experimenters. Therefore, in this section, only the practical designs for monitoring times are discussed, which are optimal for parameter estimation of the power Lindley distribution under progressive type-I interval censoring. These monitoring times cannot be fixed arbitrarily if the aim is to obtain optimal and feasible results. One of the most traditional ways is to select an equally spacing set of the monitoring times, but this choice is not suitable for decreasing failure-rate data. Meeker \cite{WQMJ} instead defined equal probability-spacing monitoring times ($\boldsymbol{t}_{\text{EP}}$) for the convenience of allowing equal proportions of failures to occur in between successive intervals. That is, using an estimated cdf $F_T(t;\hat{\alpha},\hat{\lambda})$, the quantity $u_i = i*F_T(t_r;\hat{\alpha},\hat{\lambda})/r$ is determined, where $i=1,2,...,r$ and $t_r$ is fixed beforehand. Therefore $\boldsymbol{t}_{\text{EP}} = t(\boldsymbol{u})$ is obtained, where $\boldsymbol{u}=\{u_i : i=1,2,...,r\}$. In addition, several optimal options are available for designing the monitoring times. For example, the optimal monitoring times $\boldsymbol{t}_{\text{MFI}}$ are obtained by maximizing the determinant of the expected Fisher's information matrix \cite{SS} as follows:
 \begin{equation}
 	\nonumber
 	\boldsymbol{t}_{\text{MFI}} = \arg \max\limits_{t_1 <...< t_r} [\det(\boldsymbol{I})]  
 \end{equation}
 The equally-spaced times may be used to obtain the initial parameter estimates. The latter are then utilized to maximize $\det(\boldsymbol{I})$. Another commonly approached method is to choose the monitoring times by minimizing the trace of the expected variance-covariance matrix \cite{AAI}; and these monitoring times are denoted by $\boldsymbol{t}_{\text{MA}}$:
 \begin{equation}
 	\nonumber
 	\boldsymbol{t}_{\text{MA}} = \arg \max\limits_{t_1 <...< t_r} [\text{tr}(\boldsymbol{V})]  
 \end{equation}
In this article, an E-optimality type criterion is also proposed, whereby the smallest eigenvalue of the Fisher information matrix is maximized to prevent inaccurate estimation of the parameters' directions. Denote the resulting monitoring times by $\boldsymbol{t}_{\text{ME}}$, where,
 \begin{equation}
 	\nonumber
 	\boldsymbol{t}_{\text{ME}} = \arg \max\limits_{t_1 <...< t_r} [\min \text{Eig}(\boldsymbol{I})]  
 \end{equation}
 A numerical analysis for these monitoring times is provided for a real data set in Section 5.
 \begin{table}[t!]
 	\centering
 	\caption{MSEs of the estimators for $\psi = C_p$ and $C_k$ at $r=6$.}
 	\label{tab:my-table-1}
 	\small{\begin{tabular}{cccccccccc}
 			\hline
 			\multirow{2}{*}{$\psi$}   & \multirow{2}{*}{Plan} & \multirow{2}{*}{$n$} & \multirow{2}{*}{$\hat{\psi}^{(m)}$} & \multirow{2}{*}{$\check{\psi}$} & \multirow{2}{*}{$\hat{\psi}$} & \multirow{2}{*}{$\tilde{\psi}$} & \multirow{2}{*}{$\hat{\psi}^{(b)}$} & \multirow{2}{*}{$\hat{\psi}^{(g)}$} & \multirow{2}{*}{$\hat{\psi}^{(J)}$} \\
 			& & & & & & & & & \\ \hline
 			\multirow{8}{*}{$C_p$} & \multirow{2}{*}{$\rho_1$} & 60 & \textbf{0.0157} & 0.0265 & 0.0229 & 0.0254 & 0.0200 & 0.0164 & 0.0166 \\ 
 			& & 200 & 0.0093 & 0.0082 & 0.0062 & 0.0064 & 0.0062 & \textbf{0.0057} & \textbf{0.0057} \\ \cline{2-10} 
 			& \multirow{2}{*}{$\rho_2$} & 60 & \textbf{0.0198} & 0.0421 & 0.0350 & 0.0443 & 0.0243 & 0.0224 & 0.0231 \\ 
 			& & 200 & 0.0119 & 0.0112 & 0.0091 & 0.0096 & 0.0083 & \textbf{0.0082} & 0.0083 \\ \cline{2-10} 
 			& \multirow{2}{*}{$\rho_3$}  & 60 & \textbf{0.0208} & 0.0494 & 0.0362 & 0.0479 & 0.0244 & 0.0224 & 0.0231 \\ 
 			& & 200 & 0.0140 & 0.0140 & 0.0102 & 0.0108 & 0.0096 & \textbf{0.0090} & 0.0092 \\ \cline{2-10} 
 			& \multirow{2}{*}{$\rho_4$} & 60 & \textbf{0.0186}                            & 0.0346                                 & 0.0286                               & 0.0330                                 & 0.0231                                     & 0.0194                                     & 0.0198                                     \\ 
 			&                       & 200                  & 0.0114                                     & 0.0102                                 & 0.0084                               & 0.0087                                 & 0.0081                                     & \textbf{0.0076}                            & 0.0077                                     \\ \hline
 			\multirow{8}{*}{$C_k$} & \multirow{2}{*}{$\rho_1$}    & 60                   & 0.0025                                     & 0.0030                                 & 0.0027                               & 0.0027                                 & 0.0028                                     & \textbf{0.0023}                            & \textbf{0.0023}                            \\ 
 			&                       & 200                  & 0.0014                                     & 0.0010                                 & 0.0008                               & 0.0008                                 & 0.0008                                     & \textbf{0.0007}                            & \textbf{0.0007}                            \\ \cline{2-10} 
 			& \multirow{2}{*}{$\rho_2$}   & 60                   & 0.0032                                     & 0.0039                                 & 0.0035                               & 0.0037                                 & 0.0034                                     & \textbf{0.0030}                            & \textbf{0.0030}                            \\ 
 			&                       & 200                  & 0.0018                                     & 0.0013                                 & \textbf{0.0010}                      & 0.0011                                 & \textbf{0.0010}                            & \textbf{0.0010}                            & \textbf{0.0010}                            \\ \cline{2-10} 
 			& \multirow{2}{*}{$\rho_3$}  & 60                   & 0.0034                                     & 0.0045                                 & 0.0036                               & 0.0036                                 & 0.0032                                     & \textbf{0.0030}                            & \textbf{0.0030}                            \\ 
 			&                       & 200                  & 0.0021                                     & 0.0016                                 & 0.0012                               & 0.0012                                 & \textbf{0.0011}                            & \textbf{0.0011}                            & \textbf{0.0011}                            \\ \cline{2-10} 
 			& \multirow{2}{*}{$\rho_4$}   & 60                   & 0.0030                                     & 0.0037                                 & 0.0031                               & 0.0031                                 & 0.0031                                     & \textbf{0.0027}                            & \textbf{0.0027}                            \\ 
 			&                       & 200                  & 0.0017                                     & 0.0012                                 & 0.0010                               & 0.0010                                 & 0.0010                                     & \textbf{0.0009}                            & \textbf{0.0009}                            \\ \hline
 	\end{tabular}}
 	\vspace{0.75cm}
 	\caption{MSEs of the estimators for $\psi = C_p$ and $C_k$ at $r=10$.}
 	\label{tab:my-table-2}
 	\small{\begin{tabular}{cccccccccc}
 			\hline
 			\multirow{2}{*}{$\psi$}   & \multirow{2}{*}{Plan} & \multirow{2}{*}{$n$} & \multirow{2}{*}{$\hat{\psi}^{(m)}$} & \multirow{2}{*}{$\check{\psi}$} & \multirow{2}{*}{$\hat{\psi}$} & \multirow{2}{*}{$\tilde{\psi}$} & \multirow{2}{*}{$\hat{\psi}^{(b)}$} & \multirow{2}{*}{$\hat{\psi}^{(g)}$} & \multirow{2}{*}{$\hat{\psi}^{(J)}$} \\
 			&                       &                      &                                            &                                        &                                      &                                        &                                            &                                            &                                            \\ \hline
 			\multirow{8}{*}{$C_p$} & \multirow{2}{*}{$\rho_1$}    & 60                   & \textbf{0.0124}                            & 0.0282                                 & 0.0201                               & 0.0219                                 & 0.0180                                     & 0.0147                                     & 0.0149                                     \\ 
 			&                       & 200                  & \textbf{0.0052}                            & 0.0077                                 & 0.0056                               & 0.0057                                 & 0.0056                                     & \textbf{0.0052}                            & 0.0053                                     \\ \cline{2-10} 
 			& \multirow{2}{*}{$\rho_2$}   & 60                   & \textbf{0.0156}                            & 0.0376                                 & 0.0269                               & 0.0311                                 & 0.0199                                     & 0.0196                                     & 0.0201                                     \\ 
 			&                       & 200                  & \textbf{0.0066}                            & 0.0097                                 & 0.0074                               & 0.0076                                 & 0.0071                                     & 0.0068                                     & 0.0069                                     \\ \cline{2-10} 
 			& \multirow{2}{*}{$\rho_3$}  & 60                   & \textbf{0.0147}                            & 0.0387                                 & 0.0253                               & 0.0287                                 & 0.0191                                     & 0.0180                                     & 0.0185                                     \\ 
 			&                       & 200                  & \textbf{0.0060}                            & 0.0092                                 & 0.0066                               & 0.0068                                 & 0.0060                                     & 0.0061                                     & 0.0061                                     \\ \cline{2-10} 
 			& \multirow{2}{*}{$\rho_4$}   & 60                   & \textbf{0.0148}                            & 0.0329                                 & 0.0255                               & 0.0291                                 & 0.0223                                     & 0.0179                                     & 0.0183                                     \\ 
 			&                       & 200                  & 0.0067                                     & 0.0094                                 & 0.0071                               & 0.0073                                 & 0.0070                                     & \textbf{0.0065}                            & 0.0066                                     \\ \hline
 			\multirow{8}{*}{$C_k$} & \multirow{2}{*}{$\rho_1$}    & 60                   & \textbf{0.0018}                            & 0.0029                                 & 0.0022                               & 0.0022                                 & 0.0027                                     & 0.0019                                     & 0.0019                                     \\ 
 			&                       & 200                  & 0.0007                                     & 0.0009                                 & 0.0007                               & 0.0007                                 & 0.0007                                     & \textbf{0.0006}                            & \textbf{0.0006}                            \\ \cline{2-10} 
 			& \multirow{2}{*}{$\rho_2$}   & 60                   & \textbf{0.0024}                            & 0.0037                                 & 0.0030                               & 0.0031                                 & 0.0031                                     & 0.0027                                     & 0.0026                                     \\ 
 			&                       & 200                  & 0.0010                                     & 0.0011                                 & \textbf{0.0009}                      & \textbf{0.0009}                        & 0.0010                                     & \textbf{0.0009}                            & \textbf{0.0009}                            \\ \cline{2-10} 
 			& \multirow{2}{*}{$\rho_3$}  & 60                   & \textbf{0.0022}                            & 0.0038                                 & 0.0027                               & 0.0028                                 & 0.0029                                     & 0.0024                                     & 0.0024                                     \\ 
 			&                       & 200                  & 0.0009                                     & 0.0011                                 & 0.0008                               & 0.0008                                 & 0.0008                                     & \textbf{0.0007}                            & \textbf{0.0007}                            \\ \cline{2-10} 
 			& \multirow{2}{*}{$\rho_4$}   & 60                   & \textbf{0.0023}                            & 0.0035                                 & 0.0028                               & 0.0029                                 & 0.0035                                     & 0.0025                                     & 0.0024                                     \\ 
 			&                       & 200                  & 0.0010                                     & 0.0011                                 & 0.0008                               & 0.0008                                 & 0.0009                                     & \textbf{0.0008}                            & \textbf{0.0008}                            \\ \hline
 	\end{tabular}}
 \end{table}
 \section{Simulation Studies}\label{Sec4}
 \indent In this section, the various methodologies presented in this article are analyzed using a Monte-Carlo simulation study. The behaviors of the point estimators are compared through the values of their mean squared errors (MSEs), while for the interval estimators on the basis of their coverage probabilities (CPs) and average interval lengths (AILs). All the computational work is implemented using the R programming software (version 4.5.2). Sample sizes of $n=60$ and $n=200$ are considered in two cases of monitoring times $r=6$ and $r=10$. The setting values of the parameters are $(\alpha,\lambda)=(1,0.15)$, which represents an increasing failure-rate power Lindley distribution. For ease of computation, the equally-spaced monitoring times have been arbitrarily chosen . The final monitoring time is set at $t_r = 19.053$, which is the 80$^\text{th}$ percentile of pL(1,0.15) distribution. Various plans of censoring are considered and detailed as follows:
 \begin{table}[t!]
 	\centering
 	\caption{CPs (and AILs) of the 95\% confidence and credible intervals of $\psi = C_p$ and $C_k$ at $r=6$}
 	\label{tab:my-table-3}
 	\small{
 		\begin{tabular}{ccccccccccc}
 			\hline
 			$\psi$                     & $n$                    & Plan                 & ${I}_{\text{AC}}$                                                                       & ${I}_{\text{LAC}}$                                                                      & ${I}_{\text{PB}}$                                                                       & ${I}_{\text{StB}}$                                                                       & ${I}_{\text{EBC}}(g)$                                                                      & ${I}_{\text{HPD}}(g)$                                                                       & ${I}_{\text{EBC}}(J)$                                                                      & ${I}_{\text{HPD}}(J)$                                                                       \\ \hline
 			\multirow{16}{*}{$C_p$} & \multirow{8}{*}{60}  & \multirow{2}{*}{$\rho_1$}   & \multirow{2}{*}{\begin{tabular}[c]{@{}c@{}}0.915\\ (0.5821)\end{tabular}} & \multirow{2}{*}{\begin{tabular}[c]{@{}c@{}}0.931\\ (0.5905)\end{tabular}} & \multirow{2}{*}{\begin{tabular}[c]{@{}c@{}}0.929\\ (0.6176)\end{tabular}} & \multirow{2}{*}{\begin{tabular}[c]{@{}c@{}}0.929\\ (0.6068)\end{tabular}}  & \multirow{2}{*}{\begin{tabular}[c]{@{}c@{}}\textbf{0.933}\\ (0.5175)\end{tabular}} & \multirow{2}{*}{\begin{tabular}[c]{@{}c@{}}0.931\\ (\textbf{0.5050})\end{tabular}} & \multirow{2}{*}{\begin{tabular}[c]{@{}c@{}}0.931\\ (0.5247)\end{tabular}} & \multirow{2}{*}{\begin{tabular}[c]{@{}c@{}}0.916\\ (0.5118)\end{tabular}} \\
 			&                      &                      &                                                                           &                                                                           &                                                                           &                                                                            &                                                                           &                                                                           &                                                                           &                                                                           \\  
 			&                      & \multirow{2}{*}{$\rho_2$}  & \multirow{2}{*}{\begin{tabular}[c]{@{}c@{}}0.933\\ (0.6983)\end{tabular}} & \multirow{2}{*}{\begin{tabular}[c]{@{}c@{}}0.948\\ (0.7134)\end{tabular}} & \multirow{2}{*}{\begin{tabular}[c]{@{}c@{}}0.935\\ (0.7836)\end{tabular}} & \multirow{2}{*}{\begin{tabular}[c]{@{}c@{}}0.908\\ (0.7501)\end{tabular}}  & \multirow{2}{*}{\begin{tabular}[c]{@{}c@{}}\textbf{0.949}\\ (0.6051)\end{tabular}} & \multirow{2}{*}{\begin{tabular}[c]{@{}c@{}}0.939\\ (\textbf{0.5829})\end{tabular}} & \multirow{2}{*}{\begin{tabular}[c]{@{}c@{}}0.946\\ (0.6200)\end{tabular}} & \multirow{2}{*}{\begin{tabular}[c]{@{}c@{}}0.933\\ (0.5964)\end{tabular}} \\
 			&                      &                      &                                                                           &                                                                           &                                                                           &                                                                            &                                                                           &                                                                           &                                                                           &                                                                           \\  
 			&                      & \multirow{2}{*}{$\rho_3$} & \multirow{2}{*}{\begin{tabular}[c]{@{}c@{}}0.944\\ (0.5216)\end{tabular}} & \multirow{2}{*}{\begin{tabular}[c]{@{}c@{}}\textbf{0.956}\\ (0.5341)\end{tabular}} & \multirow{2}{*}{\begin{tabular}[c]{@{}c@{}}0.949\\ (0.5908)\end{tabular}} & \multirow{2}{*}{\begin{tabular}[c]{@{}c@{}}0.921\\ (0.5547)\end{tabular}}  & \multirow{2}{*}{\begin{tabular}[c]{@{}c@{}}\textbf{0.956}\\ (0.4472)\end{tabular}} & \multirow{2}{*}{\begin{tabular}[c]{@{}c@{}}0.942\\ (\textbf{0.4303})\end{tabular}} & \multirow{2}{*}{\begin{tabular}[c]{@{}c@{}}\textbf{0.956}\\ (0.4590)\end{tabular}} & \multirow{2}{*}{\begin{tabular}[c]{@{}c@{}}0.942\\ (0.4409)\end{tabular}} \\
 			&                      &                      &                                                                           &                                                                           &                                                                           &                                                                            &                                                                           &                                                                           &                                                                           &                                                                           \\  
 			&                      & \multirow{2}{*}{$\rho_4$}  & \multirow{2}{*}{\begin{tabular}[c]{@{}c@{}}0.923\\ (0.3763)\end{tabular}} & \multirow{2}{*}{\begin{tabular}[c]{@{}c@{}}0.943\\ (0.3830)\end{tabular}} & \multirow{2}{*}{\begin{tabular}[c]{@{}c@{}}\textbf{0.945}\\ (0.4072)\end{tabular}} & \multirow{2}{*}{\begin{tabular}[c]{@{}c@{}}0.915\\ (0.3982)\end{tabular}}  & \multirow{2}{*}{\begin{tabular}[c]{@{}c@{}}\textbf{0.945}\\ (0.3290)\end{tabular}} & \multirow{2}{*}{\begin{tabular}[c]{@{}c@{}}0.932\\ (\textbf{0.3199})\end{tabular}} & \multirow{2}{*}{\begin{tabular}[c]{@{}c@{}}0.940\\ (0.3356)\end{tabular}} & \multirow{2}{*}{\begin{tabular}[c]{@{}c@{}}0.921\\ (0.3260)\end{tabular}} \\
 			&                      &                      &                                                                           &                                                                           &                                                                           &                                                                            &                                                                           &                                                                           &                                                                           &                                                                           \\ \cline{2-11}  
 			& \multirow{8}{*}{200} & \multirow{2}{*}{$\rho_1$}   & \multirow{2}{*}{\begin{tabular}[c]{@{}c@{}}0.946\\ (0.3120)\end{tabular}} & \multirow{2}{*}{\begin{tabular}[c]{@{}c@{}}0.950\\ (0.3133)\end{tabular}} & \multirow{2}{*}{\begin{tabular}[c]{@{}c@{}}0.948\\ (0.3166)\end{tabular}} & \multirow{2}{*}{\begin{tabular}[c]{@{}c@{}}0.951 \\ (0.3150)\end{tabular}} & \multirow{2}{*}{\begin{tabular}[c]{@{}c@{}}\textbf{0.958}\\ (0.3061)\end{tabular}} & \multirow{2}{*}{\begin{tabular}[c]{@{}c@{}}0.953\\ (\textbf{0.3031})\end{tabular}} & \multirow{2}{*}{\begin{tabular}[c]{@{}c@{}}0.956\\ (0.3075)\end{tabular}} & \multirow{2}{*}{\begin{tabular}[c]{@{}c@{}}0.948\\ (0.3045)\end{tabular}} \\
 			&                      &                      &                                                                           &                                                                           &                                                                           &                                                                            &                                                                           &                                                                           &                                                                           &                                                                           \\  
 			&                      & \multirow{2}{*}{$\rho_2$}  & \multirow{2}{*}{\begin{tabular}[c]{@{}c@{}}0.952\\ (0.3722)\end{tabular}} & \multirow{2}{*}{\begin{tabular}[c]{@{}c@{}}0.954\\ (0.3743)\end{tabular}} & \multirow{2}{*}{\begin{tabular}[c]{@{}c@{}}0.946\\ (0.3824)\end{tabular}} & \multirow{2}{*}{\begin{tabular}[c]{@{}c@{}}0.942\\ (0.3746)\end{tabular}}  & \multirow{2}{*}{\begin{tabular}[c]{@{}c@{}}0.952\\ (0.3648)\end{tabular}} & \multirow{2}{*}{\begin{tabular}[c]{@{}c@{}}0.950\\ (\textbf{0.3593})\end{tabular}} & \multirow{2}{*}{\begin{tabular}[c]{@{}c@{}}\textbf{0.956}\\ (0.3675)\end{tabular}} & \multirow{2}{*}{\begin{tabular}[c]{@{}c@{}}0.950\\ (0.3621)\end{tabular}} \\
 			&                      &                      &                                                                           &                                                                           &                                                                           &                                                                            &                                                                           &                                                                           &                                                                           &                                                                           \\  
 			&                      & \multirow{2}{*}{$\rho_3$} & \multirow{2}{*}{\begin{tabular}[c]{@{}c@{}}0.930\\ (0.2595)\end{tabular}} & \multirow{2}{*}{\begin{tabular}[c]{@{}c@{}}0.941\\ (0.2612)\end{tabular}} & \multirow{2}{*}{\begin{tabular}[c]{@{}c@{}}0.939\\ (0.2670)\end{tabular}} & \multirow{2}{*}{\begin{tabular}[c]{@{}c@{}}0.941\\ (0.2618)\end{tabular}}  & \multirow{2}{*}{\begin{tabular}[c]{@{}c@{}}\textbf{0.947}\\ (0.2536)\end{tabular}} & \multirow{2}{*}{\begin{tabular}[c]{@{}c@{}}0.939\\ (\textbf{0.2495})\end{tabular}} & \multirow{2}{*}{\begin{tabular}[c]{@{}c@{}}\textbf{0.947}\\ (0.2561)\end{tabular}} & \multirow{2}{*}{\begin{tabular}[c]{@{}c@{}}0.939\\ (0.2519)\end{tabular}} \\
 			&                      &                      &                                                                           &                                                                           &                                                                           &                                                                            &                                                                           &                                                                           &                                                                           &                                                                           \\  
 			&                      & \multirow{2}{*}{$\rho_4$}  & \multirow{2}{*}{\begin{tabular}[c]{@{}c@{}}0.938\\ (0.3457)\end{tabular}} & \multirow{2}{*}{\begin{tabular}[c]{@{}c@{}}0.944\\ (0.3474)\end{tabular}} & \multirow{2}{*}{\begin{tabular}[c]{@{}c@{}}0.944\\ (0.3525)\end{tabular}} & \multirow{2}{*}{\begin{tabular}[c]{@{}c@{}}0.936\\ (0.3512)\end{tabular}}  & \multirow{2}{*}{\begin{tabular}[c]{@{}c@{}}\textbf{0.952}\\ (0.3382)\end{tabular}} & \multirow{2}{*}{\begin{tabular}[c]{@{}c@{}}0.945\\ (\textbf{0.3342})\end{tabular}} & \multirow{2}{*}{\begin{tabular}[c]{@{}c@{}}0.944\\ (0.3401)\end{tabular}} & \multirow{2}{*}{\begin{tabular}[c]{@{}c@{}}0.941\\ (0.3361)\end{tabular}} \\
 			&                      &                      &                                                                           &                                                                           &                                                                           &                                                                            &                                                                           &                                                                           &                                                                           &                                                                           \\ \hline
 			\multirow{16}{*}{$C_k$} & \multirow{8}{*}{60}  & \multirow{2}{*}{$\rho_1$}   & \multirow{2}{*}{\begin{tabular}[c]{@{}c@{}}0.925\\ (0.1970)\end{tabular}} & \multirow{2}{*}{\begin{tabular}[c]{@{}c@{}}0.935\\ (0.1976)\end{tabular}} & \multirow{2}{*}{\begin{tabular}[c]{@{}c@{}}0.929\\ (0.1971)\end{tabular}} & \multirow{2}{*}{\begin{tabular}[c]{@{}c@{}}\textbf{0.975}\\ (0.1955)\end{tabular}}  & \multirow{2}{*}{\begin{tabular}[c]{@{}c@{}}0.941\\ (0.1798)\end{tabular}} & \multirow{2}{*}{\begin{tabular}[c]{@{}c@{}}0.928\\ (\textbf{0.1795})\end{tabular}} & \multirow{2}{*}{\begin{tabular}[c]{@{}c@{}}0.933\\ (0.1807)\end{tabular}} & \multirow{2}{*}{\begin{tabular}[c]{@{}c@{}}0.929\\ (0.1803)\end{tabular}} \\
 			&                      &                      &                                                                           &                                                                           &                                                                           &                                                                            &                                                                           &                                                                           &                                                                           &                                                                           \\  
 			&                      & \multirow{2}{*}{$\rho_2$}  & \multirow{2}{*}{\begin{tabular}[c]{@{}c@{}}0.923\\ (0.2311)\end{tabular}} & \multirow{2}{*}{\begin{tabular}[c]{@{}c@{}}0.940\\ (0.2321)\end{tabular}} & \multirow{2}{*}{\begin{tabular}[c]{@{}c@{}}0.935\\ (0.2319)\end{tabular}} & \multirow{2}{*}{\begin{tabular}[c]{@{}c@{}}\textbf{0.977}\\ (0.2495)\end{tabular}}  & \multirow{2}{*}{\begin{tabular}[c]{@{}c@{}}0.952\\ (0.2067)\end{tabular}} & \multirow{2}{*}{\begin{tabular}[c]{@{}c@{}}0.950\\ (\textbf{0.2063})\end{tabular}} & \multirow{2}{*}{\begin{tabular}[c]{@{}c@{}}0.946\\ (0.2084)\end{tabular}} & \multirow{2}{*}{\begin{tabular}[c]{@{}c@{}}0.942\\ (0.2079)\end{tabular}} \\
 			&                      &                      &                                                                           &                                                                           &                                                                           &                                                                            &                                                                           &                                                                           &                                                                           &                                                                           \\  
 			&                      & \multirow{2}{*}{$\rho_3$} & \multirow{2}{*}{\begin{tabular}[c]{@{}c@{}}0.944\\ (0.1720)\end{tabular}} & \multirow{2}{*}{\begin{tabular}[c]{@{}c@{}}0.947\\ (0.1729)\end{tabular}} & \multirow{2}{*}{\begin{tabular}[c]{@{}c@{}}0.949\\ (0.1713)\end{tabular}} & \multirow{2}{*}{\begin{tabular}[c]{@{}c@{}}\textbf{0.989}\\ (0.1847)\end{tabular}}  & \multirow{2}{*}{\begin{tabular}[c]{@{}c@{}}0.956\\ (0.1527)\end{tabular}} & \multirow{2}{*}{\begin{tabular}[c]{@{}c@{}}0.951\\ (\textbf{0.1524})\end{tabular}} & \multirow{2}{*}{\begin{tabular}[c]{@{}c@{}}0.956\\ (0.1539)\end{tabular}} & \multirow{2}{*}{\begin{tabular}[c]{@{}c@{}}0.949\\ (0.1536)\end{tabular}} \\
 			&                      &                      &                                                                           &                                                                           &                                                                           &                                                                            &                                                                           &                                                                           &                                                                           &                                                                           \\  
 			&                      & \multirow{2}{*}{$\rho_4$}  & \multirow{2}{*}{\begin{tabular}[c]{@{}c@{}}0.936\\ (0.1266)\end{tabular}} & \multirow{2}{*}{\begin{tabular}[c]{@{}c@{}}0.949\\ (0.1271)\end{tabular}} & \multirow{2}{*}{\begin{tabular}[c]{@{}c@{}}0.945\\ (0.1264)\end{tabular}} & \multirow{2}{*}{\begin{tabular}[c]{@{}c@{}}\textbf{0.977}\\ (0.1339)\end{tabular}}  & \multirow{2}{*}{\begin{tabular}[c]{@{}c@{}}0.938\\ (0.1144)\end{tabular}} & \multirow{2}{*}{\begin{tabular}[c]{@{}c@{}}0.932\\ (\textbf{0.1142})\end{tabular}} & \multirow{2}{*}{\begin{tabular}[c]{@{}c@{}}0.94\\ (0.1151)\end{tabular}}  & \multirow{2}{*}{\begin{tabular}[c]{@{}c@{}}0.93\\ (0.1149)\end{tabular}}  \\
 			&                      &                      &                                                                           &                                                                           &                                                                           &                                                                            &                                                                           &                                                                           &                                                                           &                                                                           \\  \cline{2-11} 
 			& \multirow{8}{*}{200} & \multirow{2}{*}{$\rho_1$}   & \multirow{2}{*}{\begin{tabular}[c]{@{}c@{}}0.946\\ (0.1083)\end{tabular}} & \multirow{2}{*}{\begin{tabular}[c]{@{}c@{}}0.942\\ (0.1084)\end{tabular}} & \multirow{2}{*}{\begin{tabular}[c]{@{}c@{}}0.948\\ (0.1083)\end{tabular}} & \multirow{2}{*}{\begin{tabular}[c]{@{}c@{}}0.959\\ (0.1093)\end{tabular}}  & \multirow{2}{*}{\begin{tabular}[c]{@{}c@{}}\textbf{0.960}\\ (0.1058)\end{tabular}} & \multirow{2}{*}{\begin{tabular}[c]{@{}c@{}}0.957\\ (\textbf{0.1057})\end{tabular}} & \multirow{2}{*}{\begin{tabular}[c]{@{}c@{}}0.956\\ (0.1060)\end{tabular}} & \multirow{2}{*}{\begin{tabular}[c]{@{}c@{}}0.955\\ (0.1058)\end{tabular}} \\
 			&                      &                      &                                                                           &                                                                           &                                                                           &                                                                            &                                                                           &                                                                           &                                                                           &                                                                           \\  
 			&                      & \multirow{2}{*}{$\rho_2$}  & \multirow{2}{*}{\begin{tabular}[c]{@{}c@{}}0.948\\ (0.1278)\end{tabular}} & \multirow{2}{*}{\begin{tabular}[c]{@{}c@{}}0.950\\ (0.1279)\end{tabular}} & \multirow{2}{*}{\begin{tabular}[c]{@{}c@{}}0.946\\ (0.1280)\end{tabular}} & \multirow{2}{*}{\begin{tabular}[c]{@{}c@{}}\textbf{0.958}\\ (0.1294)\end{tabular}}  & \multirow{2}{*}{\begin{tabular}[c]{@{}c@{}}0.952\\ (0.1240)\end{tabular}} & \multirow{2}{*}{\begin{tabular}[c]{@{}c@{}}0.952\\ (\textbf{0.1238})\end{tabular}} & \multirow{2}{*}{\begin{tabular}[c]{@{}c@{}}0.956\\ (0.1243)\end{tabular}} & \multirow{2}{*}{\begin{tabular}[c]{@{}c@{}}0.952\\ (0.1241)\end{tabular}} \\
 			&                      &                      &                                                                           &                                                                           &                                                                           &                                                                            &                                                                           &                                                                           &                                                                           &                                                                           \\  
 			&                      & \multirow{2}{*}{$\rho_3$} & \multirow{2}{*}{\begin{tabular}[c]{@{}c@{}}0.939\\ (0.0888)\end{tabular}} & \multirow{2}{*}{\begin{tabular}[c]{@{}c@{}}0.947\\ (0.0890)\end{tabular}} & \multirow{2}{*}{\begin{tabular}[c]{@{}c@{}}0.939\\ (0.0886)\end{tabular}} & \multirow{2}{*}{\begin{tabular}[c]{@{}c@{}}\textbf{0.958}\\ (0.0874)\end{tabular}}  & \multirow{2}{*}{\begin{tabular}[c]{@{}c@{}}0.947\\ (0.0860)\end{tabular}} & \multirow{2}{*}{\begin{tabular}[c]{@{}c@{}}0.946\\ (\textbf{0.0858})\end{tabular}} & \multirow{2}{*}{\begin{tabular}[c]{@{}c@{}}0.947\\ (0.0863)\end{tabular}} & \multirow{2}{*}{\begin{tabular}[c]{@{}c@{}}0.943\\ (0.0861)\end{tabular}} \\
 			&                      &                      &                                                                           &                                                                           &                                                                           &                                                                            &                                                                           &                                                                           &                                                                           &                                                                           \\  
 			&                      & \multirow{2}{*}{$\rho_4$}  & \multirow{2}{*}{\begin{tabular}[c]{@{}c@{}}0.941\\ (0.1189)\end{tabular}} & \multirow{2}{*}{\begin{tabular}[c]{@{}c@{}}0.941\\ (0.1190)\end{tabular}} & \multirow{2}{*}{\begin{tabular}[c]{@{}c@{}}0.944\\ (0.1188)\end{tabular}} & \multirow{2}{*}{\begin{tabular}[c]{@{}c@{}}\textbf{0.951}\\ (0.1206)\end{tabular}}  & \multirow{2}{*}{\begin{tabular}[c]{@{}c@{}}0.946\\ (0.1158)\end{tabular}} & \multirow{2}{*}{\begin{tabular}[c]{@{}c@{}}0.942\\ (\textbf{0.1156})\end{tabular}} & \multirow{2}{*}{\begin{tabular}[c]{@{}c@{}}0.944\\ (0.1159)\end{tabular}} & \multirow{2}{*}{\begin{tabular}[c]{@{}c@{}}0.941\\ (0.1157)\end{tabular}} \\
 			&                      &                      &                                                                           &                                                                           &                                                                           &                                                                            &                                                                           &                                                                           &                                                                           &                                                                           \\ \hline
 		\end{tabular}
 	}
 \end{table}
 \vspace{-0.35cm}
  \begin{enumerate}
 	\item $\rho_1=(0^{r-1},1)$, which is a classic type-I interval censoring scheme;
 	\item $\rho_2=(0.5,0^{r-2},1)$, where the withdrawals are considered only at the first and last monitoring times;
 	\item $\rho_3=(p^{r-1},1)$, which indicates a uniform withdrawal of survivors in each of the succeeding first $r-1$ monitoring intervals. Here, $p=0.25$ for $r=6$, and $p=0.1$ for $r=10$.
 	\item $\rho_4=(0^{\lfloor\frac{r}{2}\rfloor},0.5^{\lfloor\frac{r}{2}\rfloor-1},1)$, where the withdrawals are done only in the second half of monitoring.
 \end{enumerate}
 \vspace{-0.25cm}
  \indent To simulate the progressive type-I interval-censored data, the algorithm in Aggarwala \cite{RA} is followed. Through this algorithm, the failures $\delta_i$ within each monitoring interval $({t_{i-1},~t_i}]$, $i=1,2,...,r$ follow the conditional binomial distribution given as
 \begin{equation} \nonumber
 	\delta_i~|~\delta_{i-1},...,\delta_1,W_{i-1},...,W_1 \sim \text{Binomial}\bigg(n-\sum_{j=1}^{i-1}(\delta_j + W_j),1-\bigg[\frac{1+\lambda( t_{i}^\alpha+1)}{1 + \lambda( t_{i-1}^\alpha+1)}\bigg]e^{-\lambda(t_i^\alpha - t_{i-1}^\alpha)}\bigg)
 \end{equation}
The mid-point method is used for obtaining the initial approximations for the method of moments, maximum likelihood estimation, non-linear least squares and Bayesian methods. 
   \begin{table}[t!]
 	\centering
 	\caption{CPs (and AILs) of the 95\% confidence and credible intervals of $\psi = C_p$ and $C_k$ at $r=10$}
 	\label{tab:my-table-4}
 	\small{\begin{tabular}{ccccccccccc}
 			\hline
 			$\psi$                     & $n$                    & Plan                 & ${I}_{\text{AC}}$                                                                       & ${I}_{\text{LAC}}$                                                                      & ${I}_{\text{PB}}$                                                                       & ${I}_{\text{StB}}$                                                                      & ${I}_{\text{EBC}}(g)$                                                                      & ${I}_{\text{HPD}}(g)$                                                                       & ${I}_{\text{EBC}}(J)$                                                                      & ${I}_{\text{HPD}}(J)$                                                                       \\ \hline
 			\multirow{16}{*}{$C_p$} & \multirow{8}{*}{60}  & \multirow{2}{*}{$\rho_1$}   & \multirow{2}{*}{\begin{tabular}[c]{@{}c@{}}0.942\\ (0.5351)\end{tabular}} & \multirow{2}{*}{\begin{tabular}[c]{@{}c@{}}\textbf{0.950}\\ (0.5416)\end{tabular}} & \multirow{2}{*}{\begin{tabular}[c]{@{}c@{}}0.947\\ (0.5552)\end{tabular}} & \multirow{2}{*}{\begin{tabular}[c]{@{}c@{}}0.941\\ (0.5485)\end{tabular}} & \multirow{2}{*}{\begin{tabular}[c]{@{}c@{}}0.947\\ (0.4850)\end{tabular}} & \multirow{2}{*}{\begin{tabular}[c]{@{}c@{}}0.941\\ (\textbf{0.4747})\end{tabular}} & \multirow{2}{*}{\begin{tabular}[c]{@{}c@{}}0.949\\ (0.4910)\end{tabular}} & \multirow{2}{*}{\begin{tabular}[c]{@{}c@{}}0.937\\ (0.4804)\end{tabular}} \\
 			&                      &                      &                                                                           &                                                                           &                                                                           &                                                                           &                                                                           &                                                                           &                                                                           &                                                                           \\  
 			&                      & \multirow{2}{*}{$\rho_2$}  & \multirow{2}{*}{\begin{tabular}[c]{@{}c@{}}0.928\\ (0.6269)\end{tabular}} & \multirow{2}{*}{\begin{tabular}[c]{@{}c@{}}\textbf{0.944}\\ (0.6376)\end{tabular}} & \multirow{2}{*}{\begin{tabular}[c]{@{}c@{}}0.935\\ (0.6748)\end{tabular}} & \multirow{2}{*}{\begin{tabular}[c]{@{}c@{}}0.941\\ (0.6694)\end{tabular}} & \multirow{2}{*}{\begin{tabular}[c]{@{}c@{}}0.940\\ (0.5604)\end{tabular}} & \multirow{2}{*}{\begin{tabular}[c]{@{}c@{}}0.918\\ (\textbf{0.5431})\end{tabular}}  & \multirow{2}{*}{\begin{tabular}[c]{@{}c@{}}0.941\\ (0.5722)\end{tabular}} & \multirow{2}{*}{\begin{tabular}[c]{@{}c@{}}0.923\\ (0.5541)\end{tabular}} \\
 			&                      &                      &                                                                           &                                                                           &                                                                           &                                                                           &                                                                           &                                                                           &                                                                           &                                                                           \\  
 			&                      & \multirow{2}{*}{$\rho_3$} & \multirow{2}{*}{\begin{tabular}[c]{@{}c@{}}0.946\\ (0.6031)\end{tabular}} & \multirow{2}{*}{\begin{tabular}[c]{@{}c@{}}\textbf{0.956}\\ (0.6124)\end{tabular}} & \multirow{2}{*}{\begin{tabular}[c]{@{}c@{}}0.947\\ (0.6407)\end{tabular}} & \multirow{2}{*}{\begin{tabular}[c]{@{}c@{}}0.929\\ (0.6251)\end{tabular}} & \multirow{2}{*}{\begin{tabular}[c]{@{}c@{}}0.954\\ (0.5408)\end{tabular}} & \multirow{2}{*}{\begin{tabular}[c]{@{}c@{}}0.944\\ (\textbf{0.5260})\end{tabular}} & \multirow{2}{*}{\begin{tabular}[c]{@{}c@{}}0.955\\ (0.5501)\end{tabular}} & \multirow{2}{*}{\begin{tabular}[c]{@{}c@{}}0.943\\ (0.5346)\end{tabular}} \\
 			&                      &                      &                                                                           &                                                                           &                                                                           &                                                                           &                                                                           &                                                                           &                                                                           &                                                                           \\  
 			&                      & \multirow{2}{*}{$\rho_4$}  & \multirow{2}{*}{\begin{tabular}[c]{@{}c@{}}0.934\\ (0.6213)\end{tabular}} & \multirow{2}{*}{\begin{tabular}[c]{@{}c@{}}0.948\\ (0.6314)\end{tabular}} & \multirow{2}{*}{\begin{tabular}[c]{@{}c@{}}0.941\\ (0.6624)\end{tabular}} & \multirow{2}{*}{\begin{tabular}[c]{@{}c@{}}0.934\\ (0.6502)\end{tabular}} & \multirow{2}{*}{\begin{tabular}[c]{@{}c@{}}\textbf{0.951}\\ (0.5539)\end{tabular}} & \multirow{2}{*}{\begin{tabular}[c]{@{}c@{}}0.931\\ (\textbf{0.5390})\end{tabular}} & \multirow{2}{*}{\begin{tabular}[c]{@{}c@{}}0.947\\ (0.5631)\end{tabular}} & \multirow{2}{*}{\begin{tabular}[c]{@{}c@{}}0.927\\ (0.5474)\end{tabular}} \\
 			&                      &                      &                                                                           &                                                                           &                                                                           &                                                                           &                                                                           &                                                                           &                                                                           &                                                                           \\ \cline{2-11}  
 			& \multirow{8}{*}{200} & \multirow{2}{*}{$\rho_1$}   & \multirow{2}{*}{\begin{tabular}[c]{@{}c@{}}0.939\\ (0.2886)\end{tabular}} & \multirow{2}{*}{\begin{tabular}[c]{@{}c@{}}0.944\\ (0.2896)\end{tabular}} & \multirow{2}{*}{\begin{tabular}[c]{@{}c@{}}0.945\\ (0.2913)\end{tabular}} & \multirow{2}{*}{\begin{tabular}[c]{@{}c@{}}0.943\\ (0.2936)\end{tabular}} & \multirow{2}{*}{\begin{tabular}[c]{@{}c@{}}\textbf{0.947}\\ (0.2844)\end{tabular}} & \multirow{2}{*}{\begin{tabular}[c]{@{}c@{}}\textbf{0.947}\\ (\textbf{0.2819})\end{tabular}} & \multirow{2}{*}{\begin{tabular}[c]{@{}c@{}}\textbf{0.947}\\ (0.2853)\end{tabular}} & \multirow{2}{*}{\begin{tabular}[c]{@{}c@{}}0.941\\ (0.2828)\end{tabular}} \\
 			&                      &                      &                                                                           &                                                                           &                                                                           &                                                                           &                                                                           &                                                                           &                                                                           &                                                                           \\  
 			&                      & \multirow{2}{*}{$\rho_2$}  & \multirow{2}{*}{\begin{tabular}[c]{@{}c@{}}0.937\\ (0.3352)\end{tabular}} & \multirow{2}{*}{\begin{tabular}[c]{@{}c@{}}0.943\\ (0.3368)\end{tabular}} & \multirow{2}{*}{\begin{tabular}[c]{@{}c@{}}0.941\\ (0.3419)\end{tabular}} & \multirow{2}{*}{\begin{tabular}[c]{@{}c@{}}0.945\\ (0.3347)\end{tabular}} & \multirow{2}{*}{\begin{tabular}[c]{@{}c@{}}\textbf{0.951}\\ (0.3302)\end{tabular}} & \multirow{2}{*}{\begin{tabular}[c]{@{}c@{}}0.941\\ (\textbf{0.3262})\end{tabular}} & \multirow{2}{*}{\begin{tabular}[c]{@{}c@{}}0.945\\ (0.3323)\end{tabular}} & \multirow{2}{*}{\begin{tabular}[c]{@{}c@{}}0.939\\ (0.3282)\end{tabular}} \\
 			&                      &                      &                                                                           &                                                                           &                                                                           &                                                                           &                                                                           &                                                                           &                                                                           &                                                                           \\  
 			&                      & \multirow{2}{*}{$\rho_3$} & \multirow{2}{*}{\begin{tabular}[c]{@{}c@{}}0.953\\ (0.3293)\end{tabular}} & \multirow{2}{*}{\begin{tabular}[c]{@{}c@{}}0.956\\ (0.3308)\end{tabular}} & \multirow{2}{*}{\begin{tabular}[c]{@{}c@{}}0.956\\ (0.3343)\end{tabular}} & \multirow{2}{*}{\begin{tabular}[c]{@{}c@{}}0.954\\ (0.3261)\end{tabular}} & \multirow{2}{*}{\begin{tabular}[c]{@{}c@{}}\textbf{0.963}\\ (0.3237)\end{tabular}} & \multirow{2}{*}{\begin{tabular}[c]{@{}c@{}}0.954\\ (\textbf{0.3201})\end{tabular}} & \multirow{2}{*}{\begin{tabular}[c]{@{}c@{}}{0.958}\\ (0.3258)\end{tabular}} & \multirow{2}{*}{\begin{tabular}[c]{@{}c@{}}0.953\\ (0.3221)\end{tabular}} \\
 			&                      &                      &                                                                           &                                                                           &                                                                           &                                                                           &                                                                           &                                                                           &                                                                           &                                                                           \\  
 			&                      & \multirow{2}{*}{$\rho_4$}  & \multirow{2}{*}{\begin{tabular}[c]{@{}c@{}}0.952\\ (0.3307)\end{tabular}} & \multirow{2}{*}{\begin{tabular}[c]{@{}c@{}}{0.956}\\ (0.3322)\end{tabular}} & \multirow{2}{*}{\begin{tabular}[c]{@{}c@{}}0.949\\ (0.3359)\end{tabular}} & \multirow{2}{*}{\begin{tabular}[c]{@{}c@{}}0.943\\ (0.3142)\end{tabular}} & \multirow{2}{*}{\begin{tabular}[c]{@{}c@{}}\textbf{0.957}\\ (0.3250)\end{tabular}} & \multirow{2}{*}{\begin{tabular}[c]{@{}c@{}}0.954\\ (\textbf{0.3214})\end{tabular}} & \multirow{2}{*}{\begin{tabular}[c]{@{}c@{}}0.955\\ (0.3268)\end{tabular}} & \multirow{2}{*}{\begin{tabular}[c]{@{}c@{}}0.954\\ (0.3232)\end{tabular}} \\
 			&                      &                      &                                                                           &                                                                           &                                                                           &                                                                           &                                                                           &                                                                           &                                                                           &                                                                           \\ \hline
 			\multirow{16}{*}{$C_k$} & \multirow{8}{*}{60}  & \multirow{2}{*}{$\rho_1$}   & \multirow{2}{*}{\begin{tabular}[c]{@{}c@{}}0.951\\ (0.1822)\end{tabular}} & \multirow{2}{*}{\begin{tabular}[c]{@{}c@{}}0.957\\ (0.1827)\end{tabular}} & \multirow{2}{*}{\begin{tabular}[c]{@{}c@{}}0.947\\ (0.1823)\end{tabular}} & \multirow{2}{*}{\begin{tabular}[c]{@{}c@{}}\textbf{0.969}\\ (0.1769)\end{tabular}} & \multirow{2}{*}{\begin{tabular}[c]{@{}c@{}}0.955\\ (0.1685)\end{tabular}} & \multirow{2}{*}{\begin{tabular}[c]{@{}c@{}}0.945\\ (\textbf{0.1682})\end{tabular}} & \multirow{2}{*}{\begin{tabular}[c]{@{}c@{}}0.949\\ (0.1691)\end{tabular}} & \multirow{2}{*}{\begin{tabular}[c]{@{}c@{}}0.943\\ (0.1689)\end{tabular}} \\
 			&                      &                      &                                                                           &                                                                           &                                                                           &                                                                           &                                                                           &                                                                           &                                                                           &                                                                           \\  
 			&                      & \multirow{2}{*}{$\rho_2$}  & \multirow{2}{*}{\begin{tabular}[c]{@{}c@{}}0.940\\ (0.2112)\end{tabular}} & \multirow{2}{*}{\begin{tabular}[c]{@{}c@{}}0.955\\ (0.2120)\end{tabular}} & \multirow{2}{*}{\begin{tabular}[c]{@{}c@{}}0.935\\ (0.2136)\end{tabular}} & \multirow{2}{*}{\begin{tabular}[c]{@{}c@{}}\textbf{0.979}\\ (0.2287)\end{tabular}} & \multirow{2}{*}{\begin{tabular}[c]{@{}c@{}}0.941\\ (0.1926)\end{tabular}} & \multirow{2}{*}{\begin{tabular}[c]{@{}c@{}}0.937\\ (\textbf{0.1922})\end{tabular}} & \multirow{2}{*}{\begin{tabular}[c]{@{}c@{}}0.941\\ (0.1940)\end{tabular}} & \multirow{2}{*}{\begin{tabular}[c]{@{}c@{}}0.933\\ (0.1937)\end{tabular}} \\
 			&                      &                      &                                                                           &                                                                           &                                                                           &                                                                           &                                                                           &                                                                           &                                                                           &                                                                           \\  
 			&                      & \multirow{2}{*}{$\rho_3$} & \multirow{2}{*}{\begin{tabular}[c]{@{}c@{}}0.948\\ (0.2021)\end{tabular}} & \multirow{2}{*}{\begin{tabular}[c]{@{}c@{}}0.951\\ (0.2028)\end{tabular}} & \multirow{2}{*}{\begin{tabular}[c]{@{}c@{}}0.947\\ (0.2025)\end{tabular}} & \multirow{2}{*}{\begin{tabular}[c]{@{}c@{}}\textbf{0.966}\\ (0.2142)\end{tabular}} & \multirow{2}{*}{\begin{tabular}[c]{@{}c@{}}0.954\\ (0.1851)\end{tabular}} & \multirow{2}{*}{\begin{tabular}[c]{@{}c@{}}0.951\\ (\textbf{0.1848})\end{tabular}} & \multirow{2}{*}{\begin{tabular}[c]{@{}c@{}}0.955\\ (0.1860)\end{tabular}} & \multirow{2}{*}{\begin{tabular}[c]{@{}c@{}}0.947\\ (0.1857)\end{tabular}} \\
 			&                      &                      &                                                                           &                                                                           &                                                                           &                                                                           &                                                                           &                                                                           &                                                                           &                                                                           \\  
 			&                      & \multirow{2}{*}{$\rho_4$}  & \multirow{2}{*}{\begin{tabular}[c]{@{}c@{}}0.941\\ (0.2078)\end{tabular}} & \multirow{2}{*}{\begin{tabular}[c]{@{}c@{}}0.949\\ (0.2086)\end{tabular}} & \multirow{2}{*}{\begin{tabular}[c]{@{}c@{}}0.941\\ (0.2081)\end{tabular}} & \multirow{2}{*}{\begin{tabular}[c]{@{}c@{}}\textbf{0.974}\\ (0.2208)\end{tabular}} & \multirow{2}{*}{\begin{tabular}[c]{@{}c@{}}0.945\\ (0.1898)\end{tabular}} & \multirow{2}{*}{\begin{tabular}[c]{@{}c@{}}0.940\\ (\textbf{0.1895})\end{tabular}} & \multirow{2}{*}{\begin{tabular}[c]{@{}c@{}}0.947\\ (0.1906)\end{tabular}} & \multirow{2}{*}{\begin{tabular}[c]{@{}c@{}}0.939\\ (0.1903)\end{tabular}} \\
 			&                      &                      &                                                                           &                                                                           &                                                                           &                                                                           &                                                                           &                                                                           &                                                                           &                                                                           \\ \cline{2-11}  
 			& \multirow{8}{*}{200} & \multirow{2}{*}{$\rho_1$}   & \multirow{2}{*}{\begin{tabular}[c]{@{}c@{}}0.944\\ (0.1002)\end{tabular}} & \multirow{2}{*}{\begin{tabular}[c]{@{}c@{}}0.944\\ (0.1002)\end{tabular}} & \multirow{2}{*}{\begin{tabular}[c]{@{}c@{}}0.945\\ (0.1001)\end{tabular}} & \multirow{2}{*}{\begin{tabular}[c]{@{}c@{}}0.951\\ (0.1014)\end{tabular}} & \multirow{2}{*}{\begin{tabular}[c]{@{}c@{}}\textbf{0.952}\\ (0.0983)\end{tabular}} & \multirow{2}{*}{\begin{tabular}[c]{@{}c@{}}0.946\\ (\textbf{0.0981})\end{tabular}} & \multirow{2}{*}{\begin{tabular}[c]{@{}c@{}}0.947\\ (0.0984)\end{tabular}} & \multirow{2}{*}{\begin{tabular}[c]{@{}c@{}}0.944\\ (0.0982)\end{tabular}} \\ 
 			&                      &                      &                                                                           &                                                                           &                                                                           &                                                                           &                                                                           &                                                                           &                                                                           &                                                                           \\  
 			&                      & \multirow{2}{*}{$\rho_2$}  & \multirow{2}{*}{\begin{tabular}[c]{@{}c@{}}0.939\\ (0.1161)\end{tabular}} & \multirow{2}{*}{\begin{tabular}[c]{@{}c@{}}0.940\\ (0.1162)\end{tabular}} & \multirow{2}{*}{\begin{tabular}[c]{@{}c@{}}0.941\\ (0.1166)\end{tabular}} & \multirow{2}{*}{\begin{tabular}[c]{@{}c@{}}\textbf{0.957}\\ (0.1187)\end{tabular}} & \multirow{2}{*}{\begin{tabular}[c]{@{}c@{}}0.950\\ (0.1135)\end{tabular}} & \multirow{2}{*}{\begin{tabular}[c]{@{}c@{}}0.954\\ (\textbf{0.1133})\end{tabular}} & \multirow{2}{*}{\begin{tabular}[c]{@{}c@{}}0.945\\ (0.1138)\end{tabular}} & \multirow{2}{*}{\begin{tabular}[c]{@{}c@{}}0.941\\ (0.1136)\end{tabular}} \\
 			&                      &                      &                                                                           &                                                                           &                                                                           &                                                                           &                                                                           &                                                                           &                                                                           &                                                                           \\  
 			&                      & \multirow{2}{*}{$\rho_3$} & \multirow{2}{*}{\begin{tabular}[c]{@{}c@{}}0.956\\ (0.1134)\end{tabular}} & \multirow{2}{*}{\begin{tabular}[c]{@{}c@{}}0.961\\ (0.1136)\end{tabular}} & \multirow{2}{*}{\begin{tabular}[c]{@{}c@{}}0.956\\ (0.1134)\end{tabular}} & \multirow{2}{*}{\begin{tabular}[c]{@{}c@{}}\textbf{0.963}\\ (0.1153)\end{tabular}} & \multirow{2}{*}{\begin{tabular}[c]{@{}c@{}}0.958\\ (0.1108)\end{tabular}} & \multirow{2}{*}{\begin{tabular}[c]{@{}c@{}}0.959\\ (\textbf{0.1106})\end{tabular}} & \multirow{2}{*}{\begin{tabular}[c]{@{}c@{}}\textbf{0.963}\\ (0.1111)\end{tabular}} & \multirow{2}{*}{\begin{tabular}[c]{@{}c@{}}0.958\\ (0.1109)\end{tabular}} \\
 			&                      &                      &                                                                           &                                                                           &                                                                           &                                                                           &                                                                           &                                                                           &                                                                           &                                                                           \\  
 			&                      & \multirow{2}{*}{$\rho_4$}  & \multirow{2}{*}{\begin{tabular}[c]{@{}c@{}}0.948\\ (0.1145)\end{tabular}} & \multirow{2}{*}{\begin{tabular}[c]{@{}c@{}}0.951\\ (0.1147)\end{tabular}} & \multirow{2}{*}{\begin{tabular}[c]{@{}c@{}}0.949\\ (0.1145)\end{tabular}} & \multirow{2}{*}{\begin{tabular}[c]{@{}c@{}}0.952\\ (0.1163)\end{tabular}} & \multirow{2}{*}{\begin{tabular}[c]{@{}c@{}}\textbf{0.957}\\ (0.1119)\end{tabular}} & \multirow{2}{*}{\begin{tabular}[c]{@{}c@{}}0.955\\ (\textbf{0.1118})\end{tabular}} & \multirow{2}{*}{\begin{tabular}[c]{@{}c@{}}0.955\\ (0.1121)\end{tabular}} & \multirow{2}{*}{\begin{tabular}[c]{@{}c@{}}0.956\\ (0.1119)\end{tabular}} \\
 			&                      &                      &                                                                           &                                                                           &                                                                           &                                                                           &                                                                           &                                                                           &                                                                           &                                                                           \\ \hline
 	\end{tabular}}
 \end{table}
  1000 sets of samples are generated for each censoring plan and varying values of $n$ and $r$. Each sample is employed for all the methodologies. The overall size of each bootstrap sample is set at $S=1000$, while for the slice sampling algorithm, the length of the Markov Chain is limited to $H=45000$, with $H_b = 5000$. The MSEs of the point estimators of $C_p$ and $C_k$ are specifically computed and reported in Tables \ref{tab:my-table-1}-\ref{tab:my-table-2}. Results for the estimated parameters $\alpha$ and $\lambda$ show similar trends, for which they are excluded here. Furthermore, the interval estimators have been computed at confidence level of $\eta=0.05$. Their resulting CPs and AILs are reported in Tables \ref{tab:my-table-3}-\ref{tab:my-table-4}. The bold values in Tables \ref{tab:my-table-1}-\ref{tab:my-table-2} represent the lowest MSEs among the corresponding estimates of $C_p$ and $C_k$. It is observed that the midpoint method and the Bayesian method that employs a gamma prior substantially exhibit lower MSE values compared to the other methods. The midpoint method is efficient for small $n$, while the Bayesian method is efficient for larger $n$. Furthermore, as $n$ increases, the MSE values of all the estimators gradually decrease. As expected, an increase in $r$ further reduces the resulting MSE values. Additionally, it is observed that in all cases, the scheme $\rho_1$ produces the lowest MSE values for all methods irrespective of $n$, and $r$, and closely followed by the proposed choice $\rho_4$. \\
  \begin{table}[b!]
 	\centering
 	\caption{Comparison of fit of some lifetime models for the full stress-rupture data}
 	\label{tab:my-table-5}
 	\small{	\begin{tabular}{cccccc}
 			\hline
 			Model                                    &  $\log \mathscr{L}$                              & AIC                               & BIC   & K--S & A--D                            \\ \hline
 			Lognormal                &  -113.3079          & 230.6158         & 235.8461  & 0.1624 & 3.4171	        \\ 
 			Inverse Weibull	&-132.4394	&268.8788	&274.1091	&0.1879&	6.0901	\\
 			Inverse Gaussian	&-135.6708&	275.3416	&280.5718	&0.3168&	14.5931\\
 			Generalized Inverted Exponential&	-145.0368	&294.0736&	299.3038&	0.2596	&9.9508 \\
 			Exponential-Poisson&	-103.4497	&210.8994&	216.1296	&0.0847	&1.2332\\
 			Weibull &-102.9768&	209.9536&	215.1839&	0.0906	&1.1222 \\
 			Power Lindley&	-102.6349&	209.2699	&214.5001	&0.0833&	0.9792\\
 			\hline
 	\end{tabular}}
\end{table}
 \begin{figure}[t!]
 	\centering
 	\includegraphics[width=.425\textwidth]{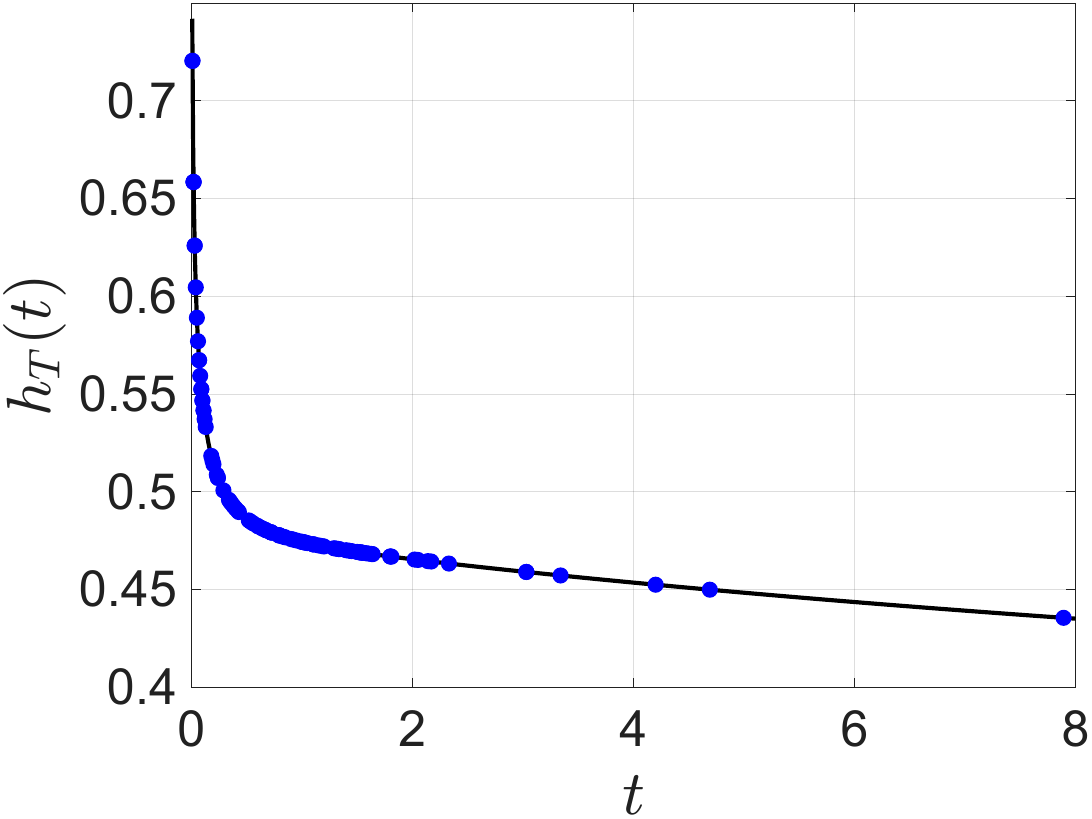}
 	\caption{Plot of the fitted failure rate function $h_T(t~|~\hat{\alpha},\hat{\lambda})$ vs $t$ using the full stress-rupture data.}
 	\label{fig1}
 \end{figure}
 \indent The bold figures in Tables \ref{tab:my-table-3}-\ref{tab:my-table-4} represent the highest CPs (and AILs) of the interval estimators. From Table \ref{tab:my-table-3}, the $I_\text{EBC}$'s derived using Gamma priors have the highest CP values for $C_p$ while the $I_\text{StB}$'s do so for $C_k$ in almost all the cases for $r=6$. However, for $r=10$, the $I_\text{LAC}$'s have the highest CP values for $C_p$, when $n=60$ in almost all cases; while $I_\text{EBC}$'s have the highest CP values for $n=200$. Similarly, the $I_\text{StB}$'s have the highest CP values for $C_k$ in almost all cases. Meanwhile, the $I_\text{HPD}$'s using Gamma priors have the shortest AILs in all cases for both $C_p$ and $C_k$, irrespective of the values of $n$ and $r$. It is also observed that for $r=10$, the scheme $\rho_1$ yields the smallest interval estimates compared to the other schemes.
 \section{Real Data Application}\label{Sec5}
\begin{figure}[t!]
 	\centering
 	\includegraphics[width=.55\textwidth]{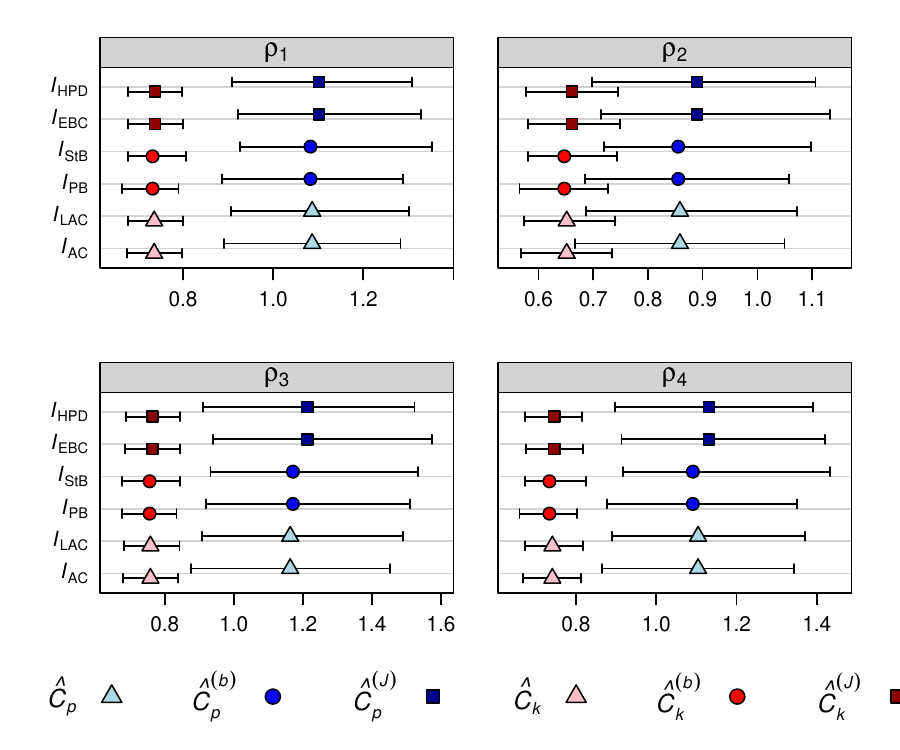}
 	\caption{The 95\% confidence and credible intervals for $C_p$ and $C_k$ along with their corresponding point estimates using full stress-rupture dataset.}
 	\label{fig2}
 	\centering
 	\includegraphics[width=0.35\textwidth]{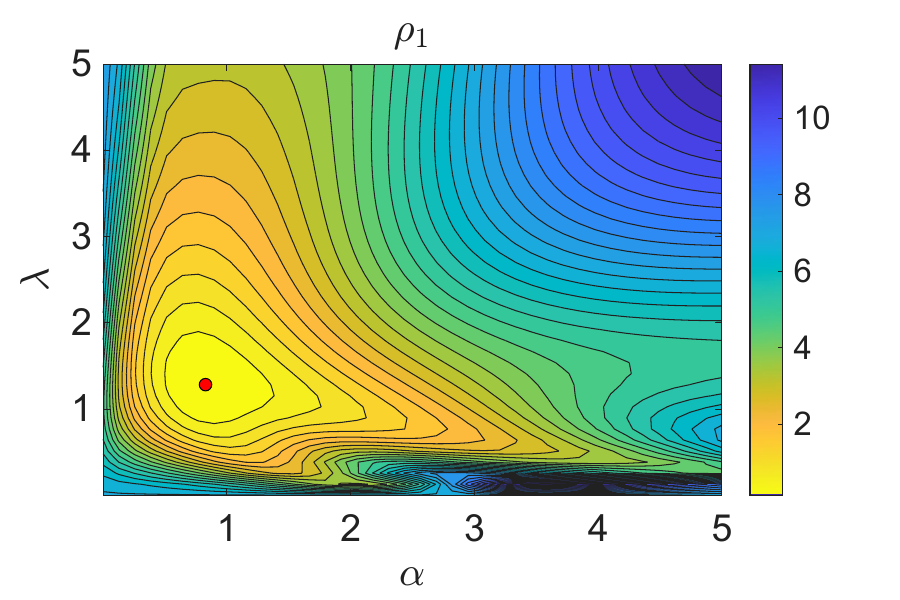}  \quad
 	\includegraphics[width=0.35\textwidth]{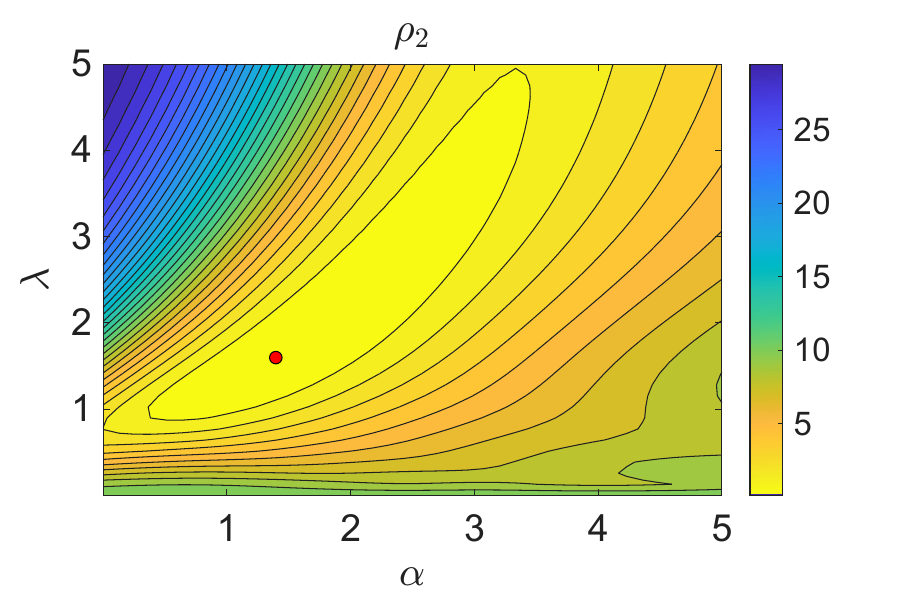} \\
 	\includegraphics[width=0.35\textwidth]{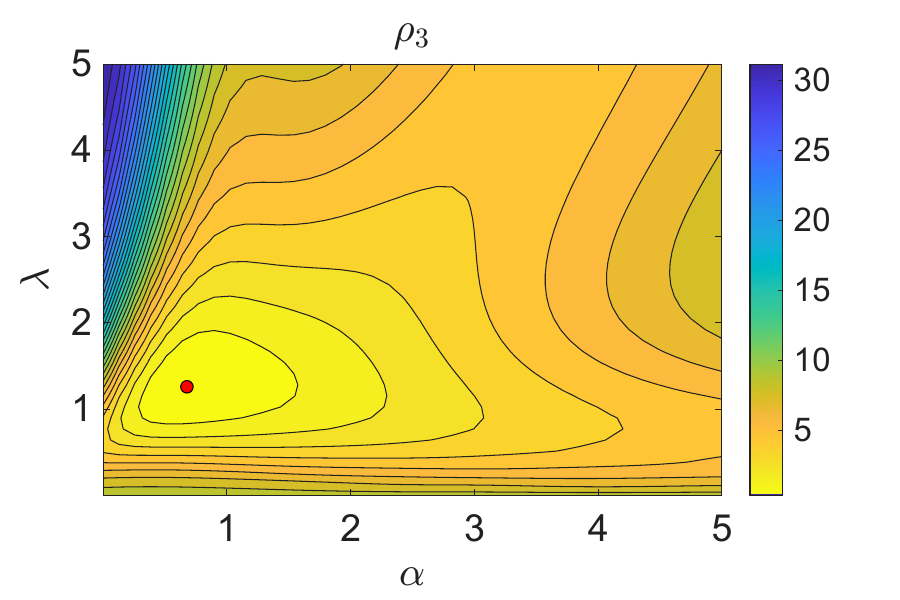}  \quad 
 	\includegraphics[width=0.35\textwidth]{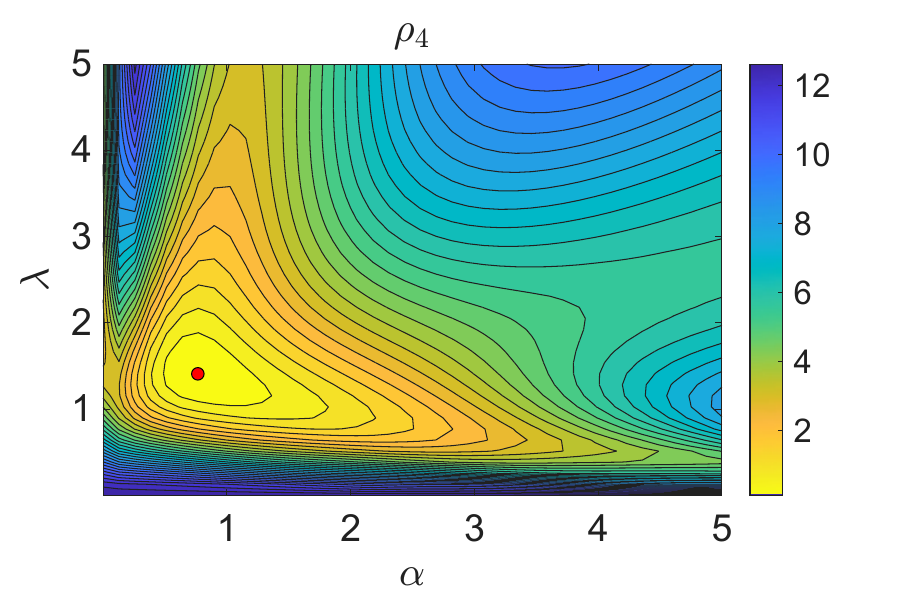} \\
 	\caption{Contour plots of the objective function $\Phi_o$ for non-linear least squares estimation for $\rho_1$ \& $\rho_2$ (Top: L-R); and $\rho_3$ \& $\rho_4$ (Bottom: L-R) for full stress-rupture dataset.}
 	\label{fig3}	
 \end{figure}

\begin{figure}[t!]
 	\includegraphics[width=0.24\textwidth]{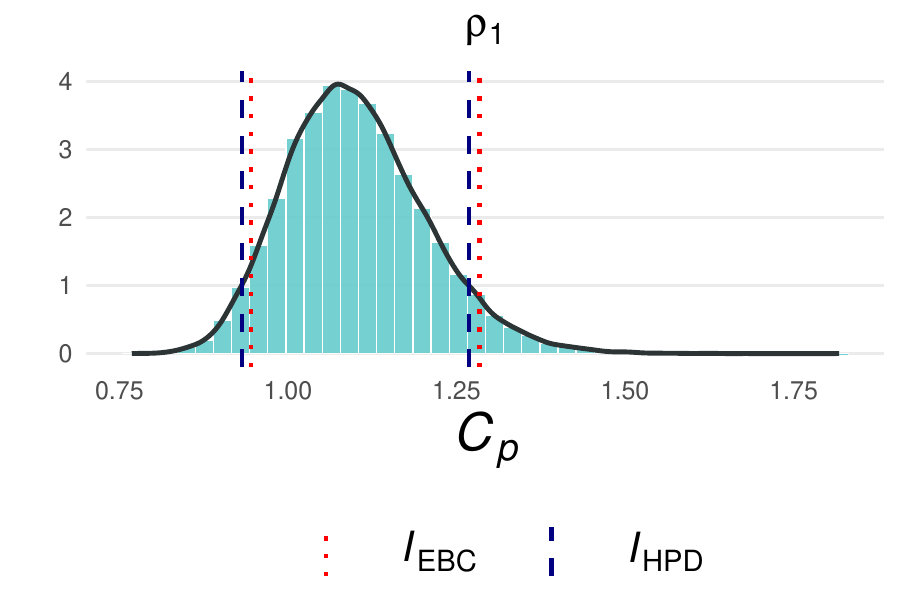}  
 	\includegraphics[width=0.24\textwidth]{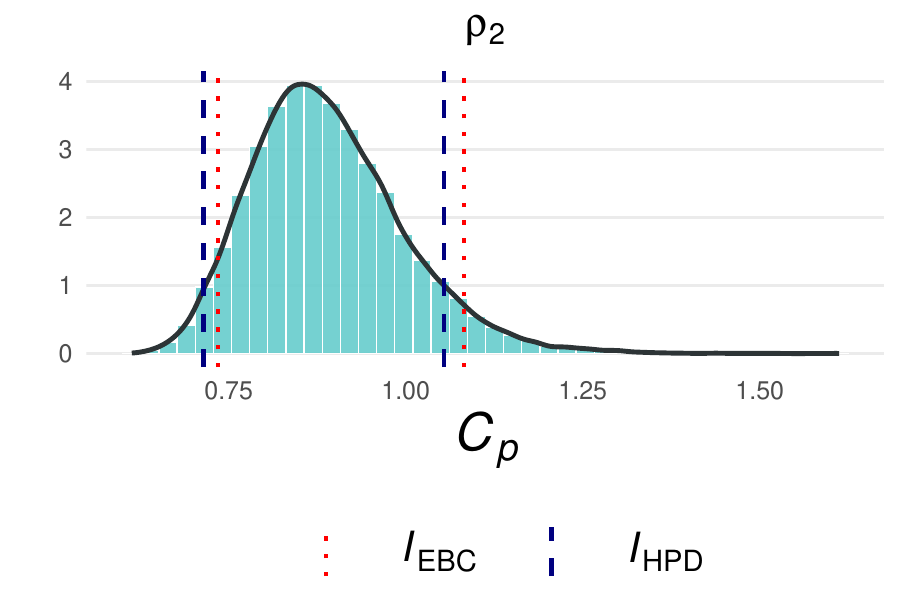}
 	\includegraphics[width=0.24\textwidth]{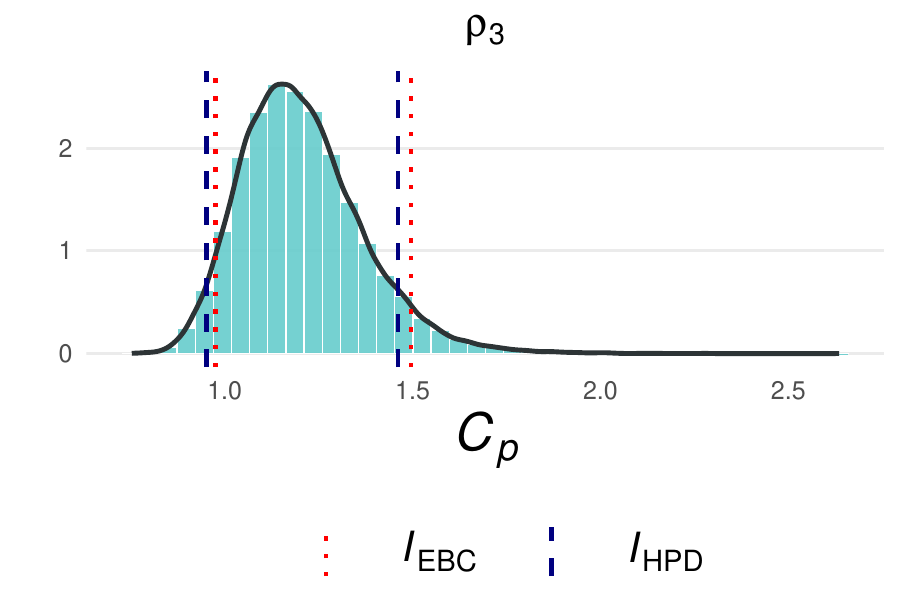}  
 	\includegraphics[width=0.24\textwidth]{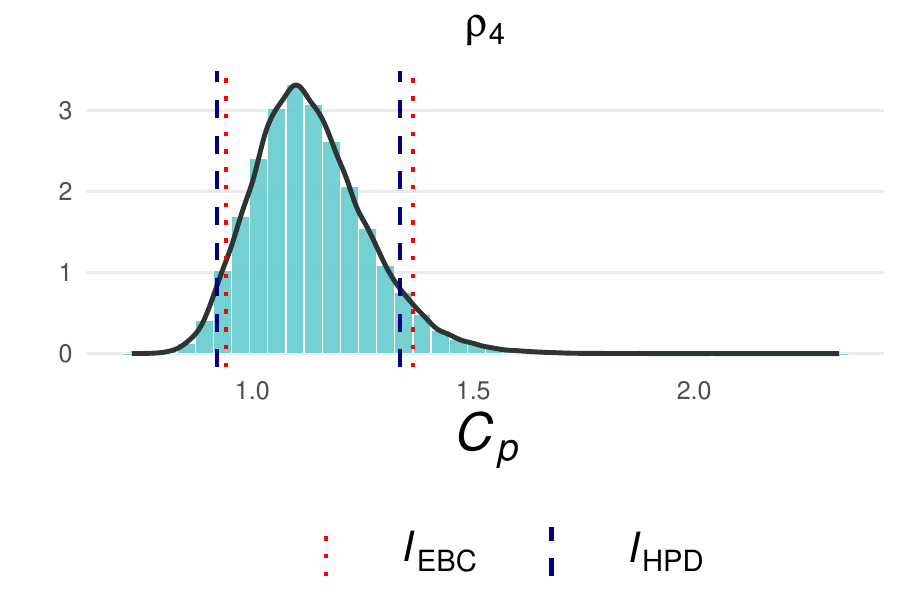} \\
 	\includegraphics[width=0.24\textwidth]{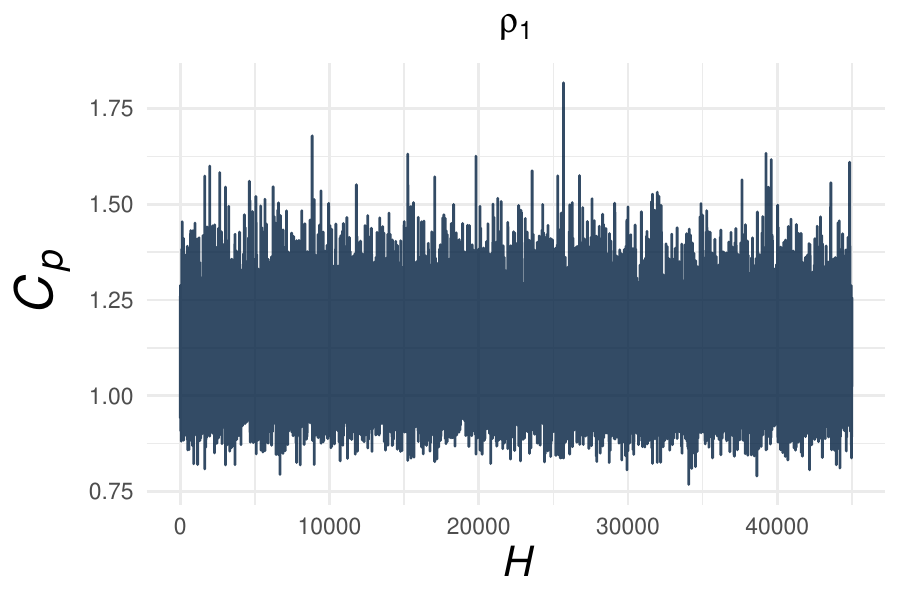}
 	\includegraphics[width=0.24\textwidth]{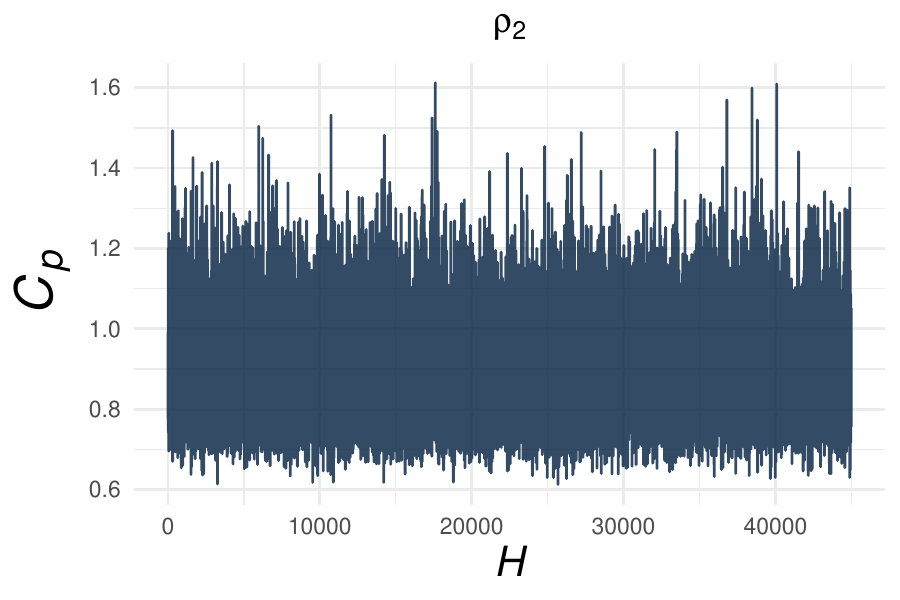} 
 	\includegraphics[width=0.24\textwidth]{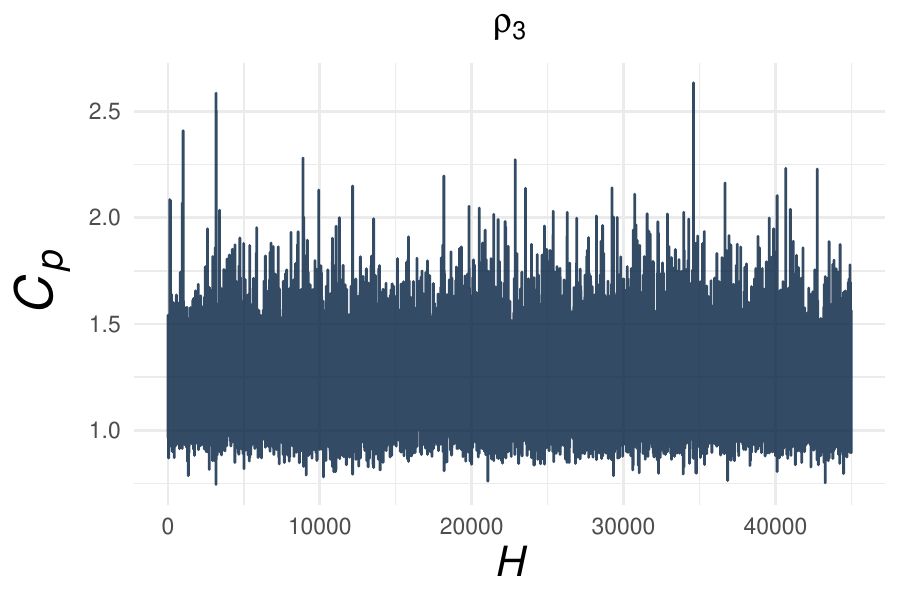} 
 	\includegraphics[width=0.24\textwidth]{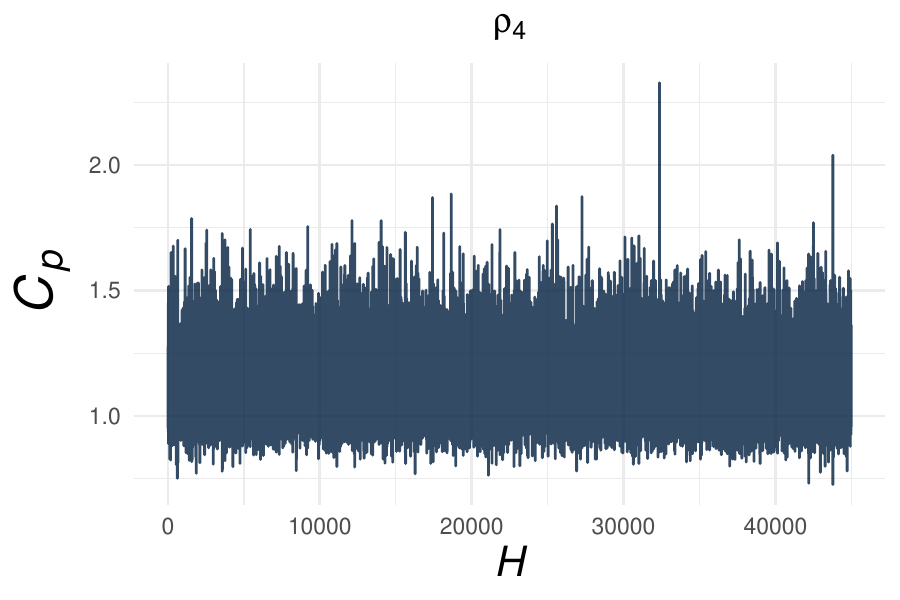} \\
 	\caption{Top (L-R): Histograms for the MCMC samples along with the ETCIs and HPDIs endpoints of $C_p$. Bottom (L-R): Traceplots of $C_p$ using full stress-rupture dataset.}
 	\label{fig4}
 	\includegraphics[width=0.24\textwidth]{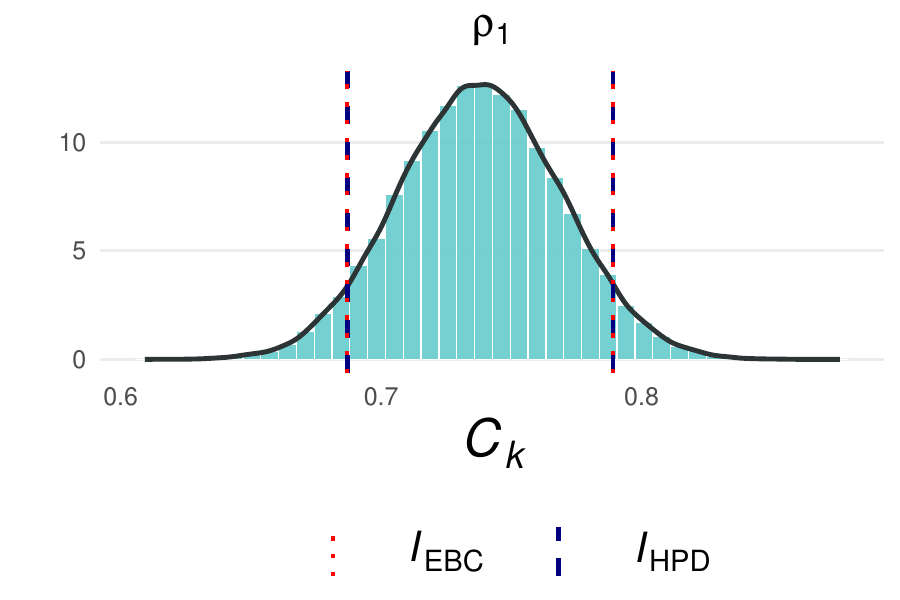}  
 	\includegraphics[width=0.24\textwidth]{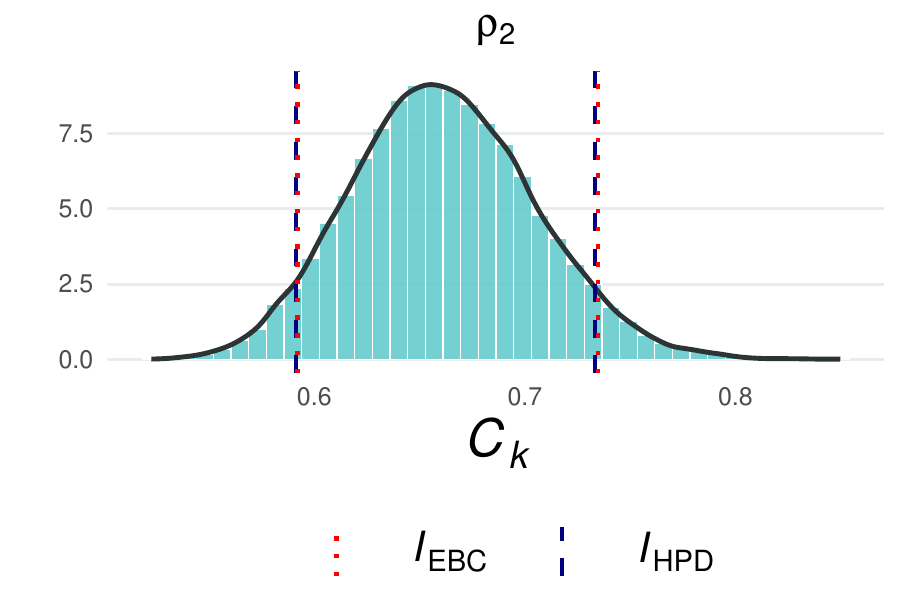}
 	\includegraphics[width=0.24\textwidth]{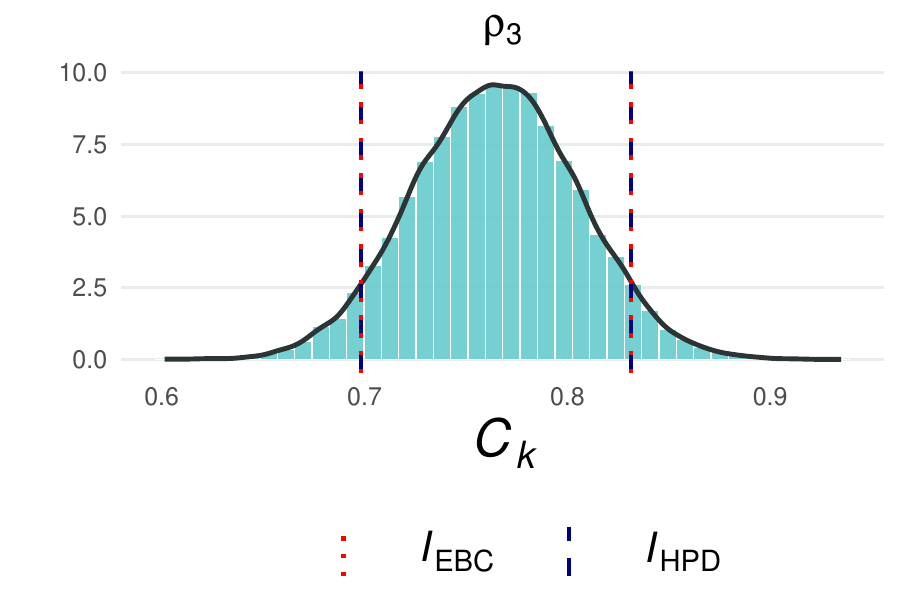}  
 	\includegraphics[width=0.24\textwidth]{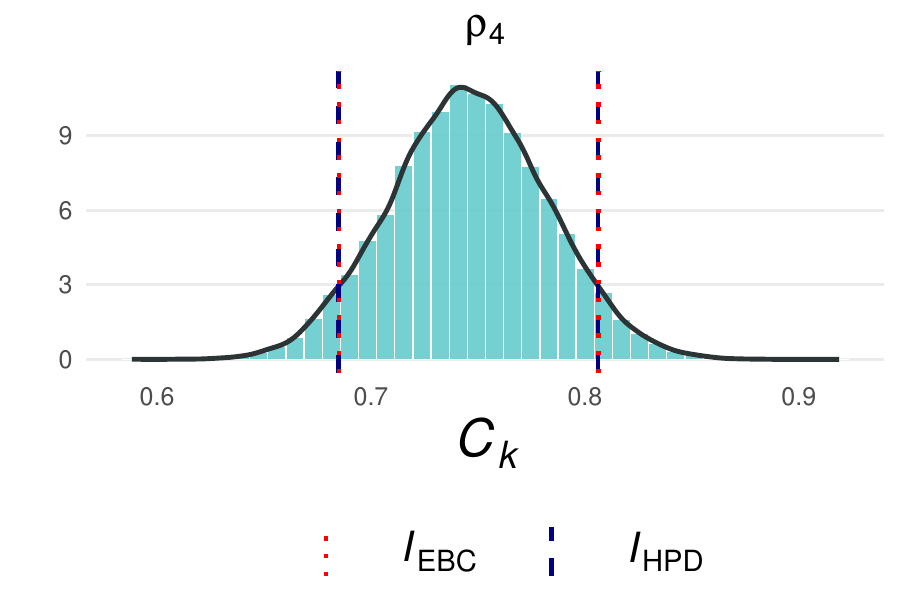} \\
 	\includegraphics[width=0.24\textwidth]{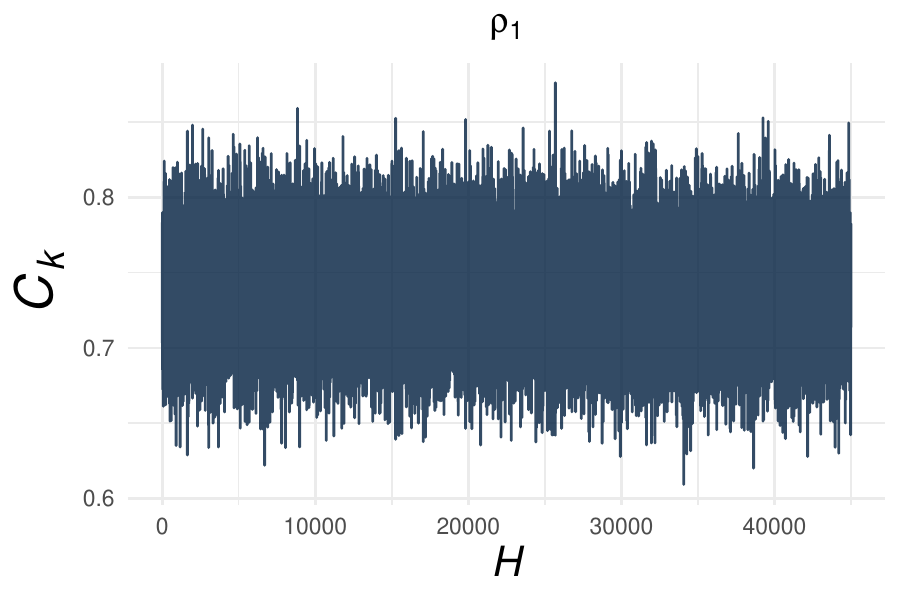}
 	\includegraphics[width=0.24\textwidth]{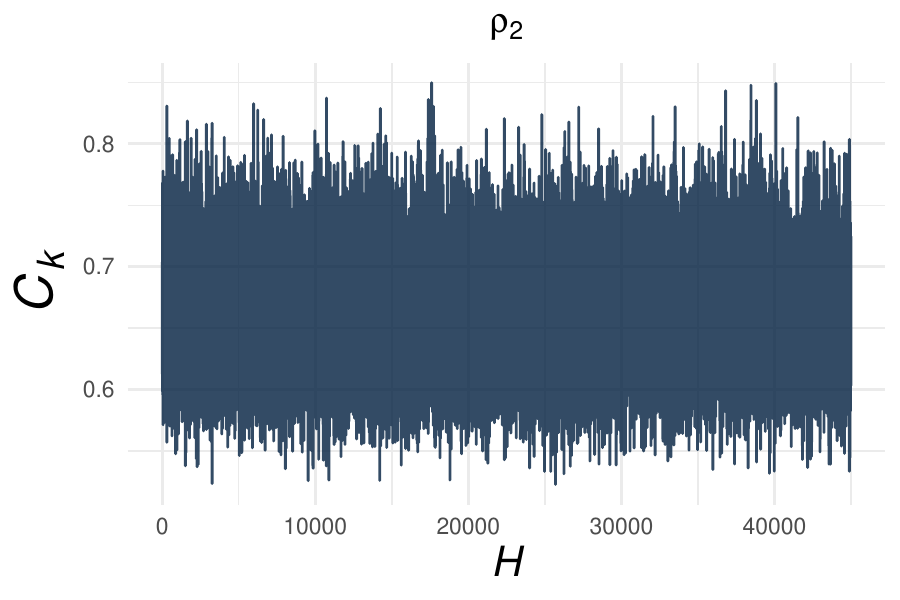} 
 	\includegraphics[width=0.24\textwidth]{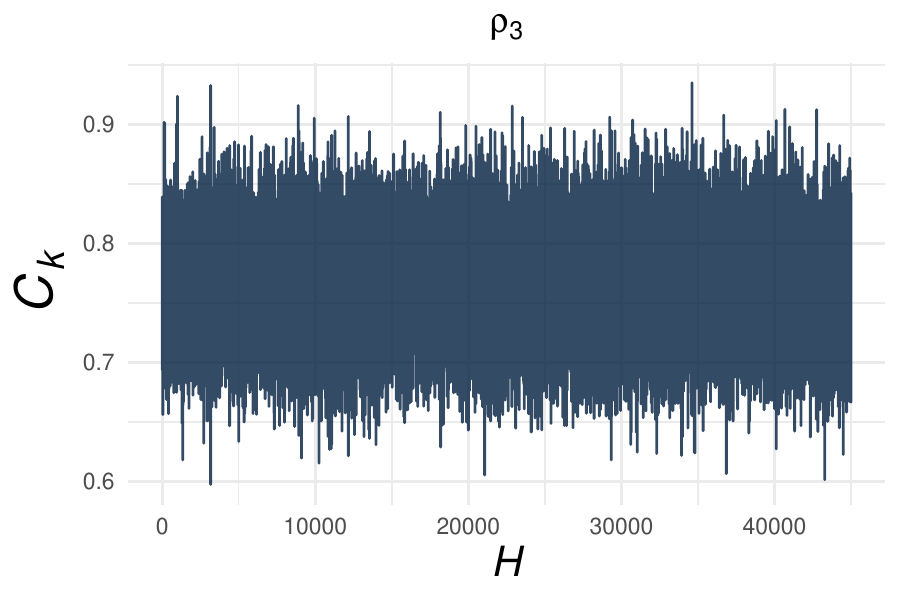} 
 	\includegraphics[width=0.24\textwidth]{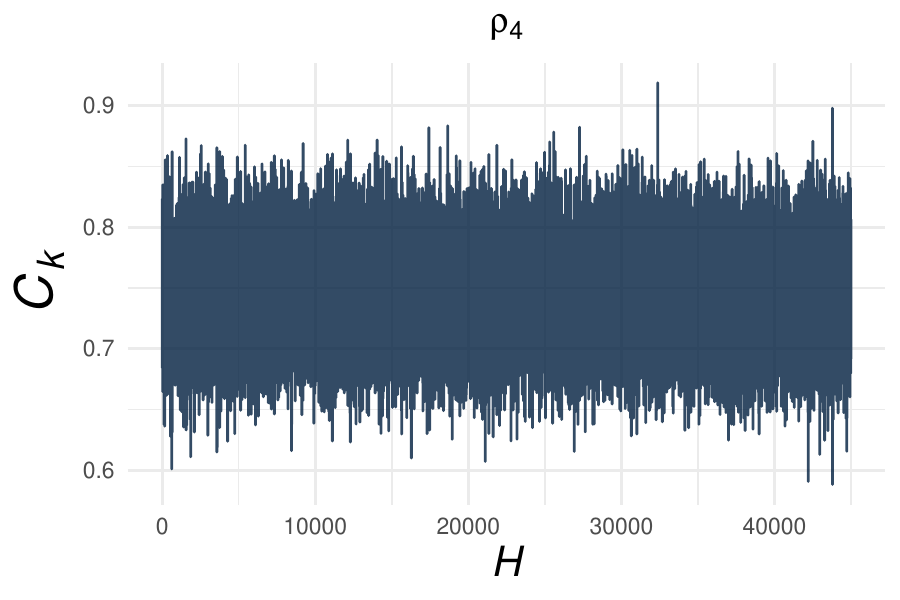} \\
 	\caption{Top (L-R): Histograms for the MCMC samples along with the ETCIs and HPDIs endpoints of $C_k$. Bottom (L-R): Traceplots of $C_k$ using full stress-rupture dataset.}
 	\label{fig5}
 \end{figure}

 \indent In this subsection, the different methods proposed in this article are applied to determine the estimates of $C_p$ and $C_k$ for a real data set presented by Kalbfleisch and Prentice \cite{JDK}. This dataset contains the full stress-rupture life of $n=101$ Kevlar 49/epoxy strands kept under constant strain at 90\% of their breaking point until all of them broke. The goodness-of-fit for the power Lindley distribution and some of the recently studied lifetime distributions under progressive type-I interval censoring--inverse Weibull \cite{SS}, inverse Gaussian \cite{SR2}, generalized inverted exponential \cite{RH}, exponential-Poisson \cite{AM} and Weibull \cite{BMN}, are evaluated through the log-likelihood value, the Akaike and Bayesian information criteria (AIC and BIC), the Kolmogorov--Smirnov (K--S) and Anderson--Darling (A--D) tests. The test results summarized in Table \ref{tab:my-table-5}. It is observed that the power Lindley model has the lowest values in terms of the fitting criteria, indicating the most optimal fit compared to the other lifetime distributions. \\
 \begin{table}[t!]
 	\centering
 	\caption{Asymptotic relative efficiencies of the monitoring times using full stress-rupture data.}
 	\label{tab:my-table-6}
 	\small{\begin{tabular}{ccccccccccc}
 			\hline
 			\multirow{2}{*}{$\rho$}   & \multirow{2}{*}{Criterion}   & \multicolumn{6}{c}{Monitoring Times}                                                                                                                  & \multirow{2}{*}{$e_{C_p}$} & \multirow{2}{*}{$e_{C_k}$} & \multirow{2}{*}{$\frac{e_{C_p} + e_{C_k}}{2}$} \\ \cline{3-8}
 			&                              & \multicolumn{1}{c}{$t_1$} & \multicolumn{1}{c}{$t_2$} & \multicolumn{1}{c}{$t_3$} & \multicolumn{1}{c}{$t_4$} & \multicolumn{1}{c}{$t_5$} & $t_6$ &                            &                            &                                                \\ \hline
 			\multirow{4}{*}{$\rho_1$} & $\boldsymbol{t}_{\text{EP}}$ & \multicolumn{1}{c}{0.146} & \multicolumn{1}{c}{0.352} & \multicolumn{1}{c}{0.623} & \multicolumn{1}{c}{0.997} & \multicolumn{1}{c}{1.589} & 3.030 & 0.748                      & 0.667                      & 0.708                                          \\  
 			& $\boldsymbol{t}_{\text{MI}}$ & \multicolumn{1}{c}{0.063} & \multicolumn{1}{c}{0.515} & \multicolumn{1}{c}{0.788} & \multicolumn{1}{c}{1.240} & \multicolumn{1}{c}{1.673} & 3.030 & 0.776                      & 0.727                      & 0.751                                          \\  
 			& $\boldsymbol{t}_{\text{ME}}$ & \multicolumn{1}{c}{0.053} & \multicolumn{1}{c}{0.335} & \multicolumn{1}{c}{0.668} & \multicolumn{1}{c}{1.140} & \multicolumn{1}{c}{1.633} & 3.030 & 0.817                      & 0.728                      & 0.772                                          \\  
 			& $\boldsymbol{t}_{\text{MA}}$ & \multicolumn{1}{c}{0.023} & \multicolumn{1}{c}{0.135} & \multicolumn{1}{c}{0.428} & \multicolumn{1}{c}{0.920} & \multicolumn{1}{c}{1.423} & 3.030 & 0.802                      & 0.743                      & 0.773                                          \\ \hline
 			\multirow{4}{*}{$\rho_2$} & $\boldsymbol{t}_{\text{EP}}$ & \multicolumn{1}{c}{0.146} & \multicolumn{1}{c}{0.352} & \multicolumn{1}{c}{0.623} & \multicolumn{1}{c}{0.997} & \multicolumn{1}{c}{1.589} & 3.030 & 0.470                      & 0.521                      & 0.495                                          \\  
 			& $\boldsymbol{t}_{\text{MI}}$ & \multicolumn{1}{c}{0.233} & \multicolumn{1}{c}{0.715} & \multicolumn{1}{c}{1.058} & \multicolumn{1}{c}{1.510} & \multicolumn{1}{c}{1.773} & 3.030 & 0.482                      & 0.537                      & 0.510                                          \\  
 			& $\boldsymbol{t}_{\text{ME}}$ & \multicolumn{1}{c}{0.503} & \multicolumn{1}{c}{0.785} & \multicolumn{1}{c}{1.178} & \multicolumn{1}{c}{1.670} & \multicolumn{1}{c}{2.123} & 3.030 & 0.723                      & 0.407                      & 0.565                                          \\  
 			& $\boldsymbol{t}_{\text{MA}}$ & \multicolumn{1}{c}{0.133} & \multicolumn{1}{c}{0.455} & \multicolumn{1}{c}{0.918} & \multicolumn{1}{c}{1.420} & \multicolumn{1}{c}{1.923} & 3.030 & 0.529                      & 0.503                      & 0.516                                          \\ \hline
 			\multirow{4}{*}{$\rho_3$} & $\boldsymbol{t}_{\text{EP}}$ & \multicolumn{1}{c}{0.146} & \multicolumn{1}{c}{0.352} & \multicolumn{1}{c}{0.623} & \multicolumn{1}{c}{0.997} & \multicolumn{1}{c}{1.589} & 3.030 & 0.304                      & 0.393                      & 0.348                                          \\  
 			& $\boldsymbol{t}_{\text{MI}}$ & \multicolumn{1}{c}{0.293} & \multicolumn{1}{c}{0.775} & \multicolumn{1}{c}{1.278} & \multicolumn{1}{c}{1.700} & \multicolumn{1}{c}{2.103} & 3.030 & 0.359                      & 0.434                      & 0.396                                          \\  
 			& $\boldsymbol{t}_{\text{ME}}$ & \multicolumn{1}{c}{0.423} & \multicolumn{1}{c}{0.875} & \multicolumn{1}{c}{1.078} & \multicolumn{1}{c}{1.240} & \multicolumn{1}{c}{1.673} & 3.030 & 0.453                      & 0.331                      & 0.392                                          \\  
 			& $\boldsymbol{t}_{\text{MA}}$ & \multicolumn{1}{c}{0.113} & \multicolumn{1}{c}{0.575} & \multicolumn{1}{c}{1.078} & \multicolumn{1}{c}{1.510} & \multicolumn{1}{c}{1.983} & 3.030 & 0.388                      & 0.438                      & 0.413                                          \\ \hline
 			\multirow{4}{*}{$\rho_4$} & $\boldsymbol{t}_{\text{EP}}$ & \multicolumn{1}{c}{0.146} & \multicolumn{1}{c}{0.352} & \multicolumn{1}{c}{0.623} & \multicolumn{1}{c}{0.997} & \multicolumn{1}{c}{1.589} & 3.030 & 0.402                      & 0.457                      & 0.430                                          \\  
 			& $\boldsymbol{t}_{\text{MI}}$ & \multicolumn{1}{c}{0.173} & \multicolumn{1}{c}{0.655} & \multicolumn{1}{c}{1.148} & \multicolumn{1}{c}{1.640} & \multicolumn{1}{c}{2.093} & 3.030 & 0.510                      & 0.524                      & 0.517                                          \\  
 			& $\boldsymbol{t}_{\text{ME}}$ & \multicolumn{1}{c}{0.083} & \multicolumn{1}{c}{0.335} & \multicolumn{1}{c}{0.738} & \multicolumn{1}{c}{1.210} & \multicolumn{1}{c}{1.563} & 3.030 & 0.441                      & 0.522                      & 0.482                                          \\  
 			& $\boldsymbol{t}_{\text{MA}}$ & \multicolumn{1}{c}{0.103} & \multicolumn{1}{c}{0.305} & \multicolumn{1}{c}{0.808} & \multicolumn{1}{c}{1.310} & \multicolumn{1}{c}{1.793} & 3.030 & 0.563                      & 0.556                      & 0.560                                          \\ \hline
 	\end{tabular}}
\vspace{0.5cm}
 	\centering
 	\caption{Progressive type-I interval censored sampling plans for the full stress-rupture dataset.}
 	\label{tab:my-table-7}
 	\small{	\begin{tabular}{clccccccc}
 			\hline
 			\multicolumn{2}{c}{Plan}                                                   & \multicolumn{1}{l}{}        & $i=1$ & $i=2$ & $i=3$ & $i=4$ & $i=5$ & $i=6$ \\ \hline
 			\multicolumn{2}{c}{\multirow{4}{*}{$\rho_1=(0,0,0,0,0,1)$}}                & $\boldsymbol{t}_{\text{MA}}$ & 0.023 & 0.135 & 0.428 & 0.920 & 1.423 & 3.030 \\ 
 			\multicolumn{2}{c}{}                                                       & $\boldsymbol{\delta}$        & 5     & 16    & 13    & 22    & 18    & 23    \\ 
 			\multicolumn{2}{c}{}                                                       & $\boldsymbol{W}$        & 0     & 0     & 0     & 0     & 0     & 4     \\ \hline
 			\multicolumn{2}{c}{\multirow{4}{*}{$\rho_2=(0.5,0,0,0,0,1)$}}              & $\boldsymbol{t}_{\text{ME}}$ & 0.503 & 0.785 & 1.178 & 1.670 & 2.123 & 3.030 \\ 
 			\multicolumn{2}{c}{}                                                       & $\boldsymbol{\delta}$        & 35    & 4     & 15    & 7     & 1     & 4     \\ 
 			\multicolumn{2}{c}{}                                                       & $\boldsymbol{W}$        & 33    & 0     & 0     & 0     & 0     & 2     \\ \hline
 			\multicolumn{2}{c}{\multirow{4}{*}{$\rho_3=(0.25,0.25,0.25,0.25,0.25,1)$}} & $\boldsymbol{t}_{\text{MA}}$ & 0.113 & 0.575 & 1.078 & 1.510 & 1.983 & 3.030 \\ 
 			\multicolumn{2}{c}{}                                                       & $\boldsymbol{\delta}$        & 19    & 14    & 15    & 5     & 5     & 2     \\ 
 			\multicolumn{2}{c}{}                                                       & $\boldsymbol{W}$        & 20    & 12    & 5     & 2     & 1     & 1     \\ \hline
 			\multicolumn{2}{c}{\multirow{4}{*}{$\rho_4=(0,0,0,0,0.5,0.5,0.5,1)$}}      & $\boldsymbol{t}_{\text{MA}}$ & 0.103 & 0.305 & 0.808 & 1.310 & 1.793 & 3.030 \\ 
 			\multicolumn{2}{c}{}                                                       & $\boldsymbol{\delta}$        & 17    & 11    & 24    & 19    & 8     & 3     \\ 
 			\multicolumn{2}{c}{}                                                       & $\boldsymbol{W}$        & 0     & 0     & 0     & 15    & 3     & 1     \\ \hline
 	\end{tabular}}
 \end{table}
 \indent Now, the type of criteria for the monitoring intervals is discussed. As evident from Figure \ref{fig1}, the data consists of several early failures, and the failure rate function is strictly decreasing. Therefore the equally spaced monitoring times are not suitable for this example\cite{WQMJ}. The real data is used to simulate 1000 cases for monitoring and compute the associated asymptotic relative efficiencies $e_\psi$ of the MLEs of $\psi=C_p$ and $C_k$, and the average of these two. Here, $e_\psi$ is defined as the ratio of the asymptotic variance estimate of the MLE of $\psi$ under complete data to that of under progressive type-I interval censoring. As an illustrative example, the same plans $\rho_1,\rho_2,\rho_3$ and $\rho_4$ discussed in the simulation study are considered in this real data analysis as well, together with $r=6$ and $t_r=3.030$, the 95$^\text{th}$ sample percentile. The results are tabulated in Table \ref{tab:my-table-6}. It is observed that the average of $e_{C_p}$ and $e_{C_k}$ is highest for $\boldsymbol{t}_{\text{ME}}$ under $\rho_2$ and $\boldsymbol{t}_{\text{MA}}$ under $\rho_1, \rho_3$ and $\rho_4$. The selection criteria can be based on either the highest $e_{C_p}$ or $e_{C_k}$ values. For the sake of illustration, the monitoring times based on the highest average of $e_{C_p}$ and $e_{C_k}$ values, are used to generate the progressive type-I interval censored data including $\boldsymbol{\delta}$ and $\boldsymbol{W}$. The latter are used to estimate the parameters and the measures $C_p$ and $C_k$ under progressive type-I interval censoring. The resulting data is given in Table \ref{tab:my-table-7}. \\
 \indent The results of point estimation (for $C_p$ and $C_k$ only) are presented in Table \ref{tab:my-table-8}, and for interval estimation in Figure \ref{fig2}. The results indicate that the estimated values are consistently close to each other for every corresponding plan. It stands to conclusion that since the $C_p$ estimates are close to unity, the full stress-rupture dataset exhibits substantial relative variability, since the standard deviation is approximately the same magnitude as the mean. \\
 \indent For $C_p$, the intervals $I_{\text{AC}}$ are smallest in the case of $\rho_1$ and $\rho_3$, while $I_{\text{PB}}$ are the smallest for $\rho_2$ and $\rho_4$. For $C_k$, the intervals $I_{\text{HPD}}$ are smallest in the case of $\rho_1$ and $\rho_4$, while $I_{\text{PB}}$ are the smallest for $\rho_2$ and $\rho_3$. For the non-linear least squares estimation problem, the contour plots of the objective function under the four censoring plans have been highlighted in Figure \ref{fig3}. These plots have been generated in MATLAB R2026a. The points in each plot represent $(\tilde{\alpha},\tilde{\lambda})$ for each censoring plan. Furthermore, the Figures \ref{fig4}-\ref{fig5} present the histograms and traceplots for the MCMC generated-samples of $C_p$ and $C_k$, respectively.
  \begin{table}[t!]
 	\centering
 	\caption{Point estimates for the various censoring plans using full stress-rupture dataset.}
 	\label{tab:my-table-8}
 	\small{\begin{tabular}{cccccccc}
 			\hline
 			$\psi$                 & Plan     & $\hat{\psi}^{(m)}$ & $\check{\psi}$ & $\hat{\psi}$ & $\hat{\psi}^{(b)}$ & $\tilde{\psi}$ & $\hat{\psi}^{(J)}$ \\ \hline
 			\multirow{4}{*}{$C_p$} & $\rho_1$ & 1.0438             & 0.9607         & 1.0866       & 1.0828             & 1.0868         & 1.1016             \\  
 			& $\rho_2$ & 0.7776             & 0.8731         & 0.8584       & 0.8553             & 0.6749         & 0.8899             \\  
 			& $\rho_3$ & 1.0106             & 0.9824         & 1.1634       & 1.1713             & 1.3383         & 1.2127             \\  
 			& $\rho_4$ & 0.9947             & 0.9657         & 1.1042       & 1.0913             & 1.1876         & 1.1317             \\ \hline
 			\multirow{4}{*}{$C_k$} & $\rho_1$ & 0.7221             & 0.6928         & 0.7358       & 0.7323             & 0.7359         & 0.7380             \\  
 			& $\rho_2$ & 0.6138             & 0.6577         & 0.6514       & 0.6470             & 0.5594         & 0.6613             \\  
 			& $\rho_3$ & 0.7108             & 0.7008         & 0.7584       & 0.7562             & 0.8011         & 0.7649             \\  
 			& $\rho_4$ & 0.7052             & 0.6947         & 0.7412       & 0.7340             & 0.7649         & 0.7458             \\ \hline
 		\end{tabular}
 	}
 \end{table}
  \section{Conclusion}
 In this article, the various methods for estimating the relative variability measures of the power Lindley distribution under progressive type-I interval censoring are explored and investigated. Various numerical and simulation techniques have been employed to alleviate the burden of approximating the point estimators. As evident from the simulation results, the slice sampling algorithm of the Bayesian paradigm gives the most efficient performance among all the methods in point and interval estimation. In particular, the Bayesian highest posterior density intervals have the shortest average length, whereas the asymptotic and percentile bootstrap intervals are smallest in many cases for the real data set illustration. Meanwhile, the equal-tailed credible intervals derived particularly using informative prior tend to have the highest coverage probability for $C_p$ while the Student-t bootstrap interval does so for $C_k$. The optimal criteria for the choice of monitoring times have been discussed, including an E-type criterion, whereby the smallest eigenvalue of the information matrix is maximized, and the same have been applied for an illustration using a real dataset. Future scope in this direction for the power Lindley distribution includes statistical inference for the stress-strength reliability, process capability indices and related hypothesis testing procedures using progressive type-I interval censored data. 
 
\noindent {\bf Funding}: No funding was received for this research.

\noindent{\bf Conflict of interest}: The authors declare that they have no conflict of interest.

\noindent{\bf Data Availability}: A real published dataset containing the full stress-rupture life of Kevlar 49/epoxy strands by Kalbfleisch and Prentice \cite{JDK} has been used for illustration of the proposed methodologies. The data has been appropriately cited in the manuscript, and is publicly available in online repositories.

\noindent {\bf Authors' contributions}: {\bf Bankitdor M. Nongrum}: Conceptualization, Data curation, Formal analysis, Investigation, Methodology, Software, Validation, Visualization, Writing – original draft, and Writing – review \& editing. {\bf Adarsha Kumar Jena}: Conceptualization, Formal analysis, Investigation, Methodology, Supervision, Validation, Visualization, Writing – original draft, and Writing – review \& editing.

\noindent{\bf Declaration of AI Use}: No AI was used in the preparation of this manuscript.
\bibliographystyle{unsrt}
\bibliography{references}
\end{document}